\documentclass{article}

\usepackage{amsmath,amssymb,amsthm}
\usepackage{graphicx} 
\usepackage{subcaption}
\usepackage{enumitem}
\usepackage{booktabs}
\usepackage{float}

\usepackage{authblk}

\usepackage{geometry}
\usepackage{microtype}
\usepackage{hyperref}
\usepackage[capitalise,nameinlink]{cleveref}

\usepackage{csquotes}
\usepackage[
backend=biber,
style=alphabetic,
sorting=ynt
]{biblatex}
\usepackage[commandnameprefix=always]{changes}
\definechangesauthor[name=Sam, color=blue]{sf}
\definechangesauthor[name=Tobias, color=red]{ts}

\DeclareMathOperator*{\argmax}{argmax}

\newcommand{\betatrue}{\hat{\beta}_{\mathrm{TRUE}}}

\usepackage[toc,page]{appendix}
\crefname{section}{Appendix}{Appendices}   
\Crefname{section}{Appendix}{Appendices}   

\title{Principled Frequentist Estimation of Racial Disparity in Credit Approval under Unobserved Race}

\author[1]{Sam Fisher}
\author[1]{Dmitry Lesnik}
\author[1,2,3]{Tobias Sch{\"a}fer}
\affil[1]{Stratyfy Inc., New York, New York, USA}
\affil[2]{Department of Mathematics, College of Staten Island, Staten Island, NY, USA} 
\affil[3]{Physics Program, CUNY Graduate Center, NY, USA}

\begin{document}

\maketitle

\begin{abstract}
Estimating racial disparities in loan-approval probabilities when race is unobserved is routinely required for fair lending compliance. In such cases, race probabilities—typically from Bayesian Improved Surname Geocoding (BISG)—stand in for true race. \citeauthor{chen:2019} showed that common heuristic approaches, including the Threshold and Weighting estimators, are inconsistent under valid identification assumptions, compromising internal validity~\cite{chen:2019}. \citeauthor{mccartan:2024} proposed a Bayesian remedy under assumptions reasonable in many fair lending contexts~\cite{mccartan:2024}. Their approach hinges on the insight that identification requires the race predictors to be exogenous with respect to loan approval, essentially an instrumental-variables design. We present a frequentist counterpart to their solution via Ordinary Least Squares (OLS) and Maximum Likelihood Estimation (MLE) under a similar exogeneity assumption. To satisfy these assumptions in practice, we introduce (i) a surname-only proxy analogous to BISG and (ii) an income-stratified prior for race probabilities. Monte Carlo simulations and an application to 2023 Los Angeles HMDA data confirm superior performance: this method reduces RMSE in the LA Black/White adverse-impact ratio by $79.7\%$ (from $10.639$\,pp to $2.158$\,pp) compared to a Weighting estimator with the standard prior.
\end{abstract}

\section*{Introduction}

Equal opportunity to access credit regardless of race and other protected attributes is a cornerstone of U.S. law under the Equal Credit Opportunity Act (ECOA) and Regulation B. This statute and its regulation not only prohibit lending discrimination on the basis of race, but also restrict creditors from inquiring about an applicant’s race with a few exceptions including home mortgage transactions covered under the Home Mortgage Disclosure Act (HMDA) and Regulation C~\cite[~p.4]{cfpb:2014}. Interagency Fair Lending Examination Guidelines call for examiners to give consideration to the use of surrogate methods to measure gender and racial approval rate disparities as an indicator of potential disparate treatment for non-HMDA reporters~\cite[~p.8-9]{ffiec:2021}. 
\\\\
In 2014, the Consumer Financial Protection Bureau published an assessment of the Bayesian Improved Surname Geocoding (BISG) methodology is use by its Office of Research and Division of
Supervision, Enforcement, and Fair Lending~\cite{cfpb:2014}. This method, introduced by \citeauthor{elliott:2009}, uses the loan applicant's surname and geo-coded residential address in conjunction with census data to derive the probability of race conditional on surname and geography~\cite{elliott:2009}. In doing so, it assumes independence of geography and surname to allow a multiplicative Bayesian update. Greengard and Gelman recently published an analysis of the statistical bias introduced by this independence assumption and proposed an improvement to BISG using joint frequency tables of surname and geolocation available in state-level voter files~\cite{greengard:2024}.
\\\\
While the CFPB report did not include information regarding its internal methods used to estimate disparities from continuous probabilities, \citeauthor{zhang:2018} introduced the assumption that the race probabilities were being used directly as regressors in a linear regression model in a manner similar to our surrogate OLS~\cite{zhang:2018}. Another technique used in practice to estimate disparity under continuous probabilities is the Weighting estimator. \citeauthor{zhang:2018} also introduced to the fair lending literature the MAP (Maximum a posteriori) method for classification of race probabilities as an alternative to the Threshold classification method. \citeauthor{zhang:2018} utilized this technique particularly for estimating pricing disparities, holding other factors equal. When no control variables are introduced in the OLS regression, this approach reduces to the MAP estimator studied in this paper via equivalence of an OLS fixed-effect coefficient to the within-group sample average. A point of semantic confusion might arise here in the use of the term MAP. In \citeauthor{mccartan:2024}'s BIRDiE (Bayesian Instrumental Regression for Disparity Estimation) family of estimators, the maximum a posteriori probability over the posterior distribution of approval rates gives a point estimate for the acceptance rate. In contrast, the ``MAP estimator" defined in this paper, per \citeauthor{zhang:2018}'s terminology, is a classification rule which takes the maximum a posteriori race probability for each loan applicant and then computes a within-group sample average of loan approval using the binarized probabilities.
\\\\
 \citeauthor{chen:2019} provided mathematical derivations for the statistical bias of disparity estimation under the Weighting and Threshold methods. The Weighting estimator was found to systematically underestimate disparity wherever the residual of predicted race is correlated with the outcome, whereas the sign of the Threshold estimator's bias depends on the threshold chosen~\cite{chen:2019}. These findings demonstrated that the Threshold and Weighting estimators common in fair lending analysis lacked the statistical property of consistency, defined as asymptotic convergence to the true disparity as the amount of sampled data increases, under realistic assumptions. Consistency is broadly considered a requirement for internal validity of an econometric analysis~\cite{stockwatson:2015}. In the context of fair lending analysis on large data samples, consistency translates to a requirement for neither over-estimating nor under-estimating disparities systematically, thus neither over-detecting nor under-detecting the presence of practically significant disparity.
 \\\\
 The 2023 analysis of \citeauthor{greenwald:2023} highlights the practical importance of the statistical bias of the Weighting estimator when used in conjunction with BISG. The Weighting method under BISG compared to an alternative Image-based race probability gave an approval rate disparity underestimate of 43\% (2.3pp vs 1.4pp) in 2017-2019 small business loan application data from online lending marketplace Lendio~\cite{greenwald:2023}. Whereas \citeauthor{greenwald:2023} focused on the improvement of the race prediction as a means to reduce bias under the Weighting estimator, we focus here on establishing consistent estimation by replacing the Weighting estimator with an alternative method. Techniques for improved race prediction include Zest Race Predictor, \citeauthor{argyle:2024}'s machine learning method incorporating additional neighborhood and individual features, and \citeauthor{imai:2022}'s fBISG (\cite{zestai:2020}, \cite{argyle:2024}, \cite{imai:2022}). \citeauthor{decterfrain:2022}'s empirical comparison of machine learning methods and BISG highlighted the heterogeneity of race classification performance across geographic regions \cite{decterfrain:2022}.
\\\\
 \citeauthor{mccartan:2024} responded to the need for a BISG-based disparity estimation methodology which exhibited consistency under realistic identifying assumptions with their 2024 paper ``Estimating Racial Disparities When Race is Not Observed" which introduced the BIRDiE (Bayesian Instrumental Regression for Disparity Estimation) methodology and an accompanying R package for implementation. The BIRDiE method effectively uses Surname as an instrument for Race and thus relies on the exogeneity assumption that surname is conditionally independent of the approval outcome given race and the other race predictors~\cite{mccartan:2024}. They call this (CI-YS), which we generalize below to (CI-YZ), where Z represents the set of variables used as predictive proxies for race. They also define an accuracy assumption, (ACC), which establishes well-calibrated posterior race probabilities that match the target population. We build our method on the same assumption, but express it in slightly different mathematical terms. Under these and other assumptions, \citeauthor{mccartan:2024} construct a family of theoretically sound Bayesian models and offer computationally efficient methods for fitting them.
 \\\\
 The formal assumptions and theoretical guarantees for the surrogate OLS method are a primary contribution of this paper building on \citeauthor{mccartan:2024}'s work. However, the bias of a similar method for estimating disparities was previously tested in a simulation study by \citeauthor{pace:2021}~\cite{pace:2021}. \citeauthor{pace:2021} studies a similar surrogate OLS (under the name BISG Continuous introduced by \citeauthor{zhang:2018}) in contrasting Disparate Treatment and Disparate Impact conditions. Under Disparate Treatment a fee is added directly based on race, such that the exogeneity condition (CI-YZ) holds and the estimator of fee disparity appears unbiased. Under Disparate Impact, a fee is added based on median income within the applicant's census block group; effectively adding a direct effect of geography on the fee outcome and breaking exogeneity. Pace observes that the OLS estimator appears materially biased in the latter scenario. We suggest that this observed result may be due to the violation of the (CI-YZ) assumption defined in this paper. Two approaches for handling a known geographic exogeneity violation are (i) BIRDiE's use of partial pooling/post-stratification and (ii) a surname-only simplification of BISG, Bayes Rule on Surname (BRS), which remains consistent at the cost of higher estimation variance.
\\\\
  Under an assumed match between census-based race statistics and the target population (ACC), the exogeneity (CI-YZ) assumption, and a few additional regularity conditions we establish the existence of a valid surrogate model for the data generating process which yields both an Ordinary Least Squares/Generalized Method of Moments (OLS, GMM) estimator as well as a Maximum Likelihood Estimator (MLE) which exhibit consistency and other desirable properties. Our proofs argue that our stated assumptions fulfill the standard conditions for both of these extremum estimators established in existing econometric theory as expressed by \citeauthor{hayashi:2000}~\cite{hayashi:2000}. The simplest and most important theoretical result is the Proof of Correct Specification and Identification which shows that the surrogate model correctly identifies the true race-conditional acceptance rates. Further, this proof indicates exactly why identification in the surrogate model hinges on the stated assumptions. 
 \\\\
 In addition to theoretical proofs, we provide a simulation study and semi-synthetic HMDA study. The simulation study (i) confirms the expected convergence behavior under a range of levels of entropy in the posterior race probabilities and (ii) illustrates the effect of violating the (CI-YZ) assumption. The semi-synthetic study on 2023 Los Angeles Home Mortgage Disclosure Act Data illustrates error reduction when replacing a heuristic approach with a standard BISG prior with a principled method under a prior better matched to the target population~\cite{ffiec:hmda2023}.

\section*{Notation}

We use the following notation throughout the paper.

\begin{align*}
Y_i &\in \{0,1\} \quad &\text{Loan approval outcome for individual } i \\
R_i &\in\ \mathcal{R} &\text{ Unobserved true race of individual } i  \\
\mathrm{T} &\in \{0,1\}^{n \times |\mathcal{R}|} \quad &\text{Matrix of one-hot true race labels (unobserved)} \\
t_i &\in \{0,1\}^{|\mathcal{R}|} \quad &\text{True race vector for individual } i \\
\Phi &= (\mathbb{P}(R=r|Z_i))^{ n \times \mathcal{R}} \quad &\text{ Matrix of posterior race probability vectors } \\
\phi_i &= (\mathbb{P}(R=r|Z_i))^{r \in \mathcal{R}} \quad &\text{ Posterior race probability vector } \\
Z_i &= \text{a set of random variables} \quad &\text{ An unspecified conditioning set } \\
S_i &\in \mathcal{S} &\text{Observed surname} \\
G_i &\in \mathcal{G} &\text{Observed geography (e.g. census tract)} \\
\beta &= \bigl(\mathbb{P}(Y_i=1 \mid R_i = r)\bigr)^{r \in \mathcal{R}}.
 \quad &\text{Race-conditional approval rate vector} \\
\Theta &= [0,1]^{|\mathcal{R}|} \quad &\text{The parameter space for the surrogate model} \\
\theta &\in \Theta \quad &\text{A point in the parameter space}
\end{align*}

\section*{Heuristic and Principled Estimators}

We introduce here three \emph{heuristic estimators} and the three \emph{principled estimators} which are the subject of the paper. By \emph{principled}, we mean estimators that emerge as solutions to well-defined optimization problems targeting the true parameter and could satisfy consistency under contextually reasonable assumptions. By \emph{heuristic}, we mean a method that does not solve a clear optimization criterion or doesn't offer a consistency guarantee under reasonable assumptions. The theoretical work of this paper is to justify two of the three principled estimators (Ordinary Least Squares and Maximum Likelihood Estimation), with the Bayesian method BIRDiE already having been developed and justified by McCartan et al \cite{mccartan:2024}. The simulation and semi-synthetic HMDA study illustrate differences in estimation error resulting from the choice of method.
\medskip
\noindent
Let $n$ denote the number of loan applications and $|\mathcal{R}|$ the number of mutually--exclusive race categories.  For the ordered set of applications $i=1,\dots,n$ we observe the binary approval outcomes
$$
  Y=(Y_1,\dots,Y_n)^{\!\top}\in\{0,1\}^{n}
$$
and the posterior race-probability matrix
$$
  \Phi=(\phi_{ir})_{i=1,\dots,n}^{r\in\mathcal{R}}
  \in[0,1]^{\,n\times|\mathcal{R}|},
  \qquad \text{where} \qquad
  \forall i. \sum_{r\in \mathcal{R}} \phi_i = 1.$$
Our target for statistical inference is the unobserved vector of race-conditional approval probabilities

$$
  \beta=(\beta_r)_{r\in\mathcal{R}},
  \qquad
  \beta_r:=\mathbb{P}(Y_i=1\mid R_i=r).
$$
Below we define the heuristic estimators we have most often encountered in practice, as well as the principled estimators.

\paragraph{H1.\ Maximum a posteriori (MAP)}  Assign each applicant to the single race with largest posterior probability  mass
$$
  \tilde t_{ir}=\mathbf 1\!\bigl\{r=\argmax_{r'\in\mathcal R}\phi_{ir'}\bigr\},
$$
and compute a pseudo race-conditional mean
$$
  \widehat\beta^{\text{MAP}}_{r}=\frac{\sum_{i=1}^{n}Y_i\tilde t_{ir}}{\sum_{i=1}^{n}\tilde t_{ir}}.
$$
We call it a pseudo-mean because it estimates the parameter under a mis-specified model where true race is replaced with the estimated race. Both the Threshold estimator and MAP method, because they use a race classification which is then used to compute a sample average, are consistent only when the pseudo-mean under the predicted race is equal to the mean under true race \cite{chen:2019}. Chen et al discuss this in the context of the Threshold estimator, but their Theorem 3.3 is applicable to any approach which uses classification to replace the true race label and then compute the sample mean within each race group. For this reason we label them both as heuristic estimators.

\paragraph{H2.\ Threshold Estimator}  Choose a cut--off $\tau\in(0,1)$ (e.g.$\tau=0.50$), set
$$
  \tilde t_{ir}=\mathbf{1}\{\phi_{ir}\ge\tau\},\qquad
  \text{drop rows with }\sum_{r}\tilde t_{ir}=0,
$$
and compute the pseudo race-conditional mean  $$\widehat\beta^{\tau}_{r}=\frac{\sum_{i=1}^{n}Y_i\tilde t_{ir}}{\sum_{i=1}^{n}\tilde t_{ir}}.$$
As with MAP, the Threshold estimator is consistent only when the pseudo-mean under predicted race is the same as the true mean. This would be true, for instance, when true race is exactly equal to predicted race. Misclassification of race can introduce bias in the estimate.

\paragraph{H3.\ Weighting Estimator}  Treat posterior probabilities as fractional race labels, partially assigning each loan approval outcome to each race using a weighted sum. Normalize the sum by the total probability mass for each race. This offers a way to generalize the sample average to the probabilistic case and get an intuitive plug-in estimate for the population mean.
$$
  \widehat\beta^{\text{WA}}_{r}=\frac{\sum_{i=1}^{n}Y_i\phi_{ir}}{\sum_{i=1}^{n}\phi_{ir}}.
$$
However, the weighted average estimator is only consistent when race $R$ is conditionally independent of loan approval $Y$ given the predictors of race $Z$ \cite{chen:2019}:
\[
  Y \;\perp\!\!\!\perp\; R \mid Z.
\]
This is an unsound assumption in most fair lending contexts because race is correlated with loan approval in many ways, particularly economic differences, not fully accounted for by surname and geography. While the Weighting estimator can be formulated as a plug-in estimator targeting the population mean, we label it as a heuristic because the independence assumption needed to do so is not reasonably satisfiable here.

\paragraph{P1.\ Ordinary Least Squares (OLS)}  Replace the unobserved one--hot race vector $t_i$ in the linear probability model $Y_i=t_i^{\!\top}\beta+\varepsilon_i$ by the posterior race probabilities $\phi_i$ and solve
$$
  \widehat\beta_{\text{OLS}}=(\Phi^{\!\top}\Phi)^{-1}\Phi^{\!\top}Y.
$$
Under the assumptions (ACC), (CI-YZ), (ID), (REG) described in the Assumptions section, this estimator is finite-sample unbiased, consistent, and asymptotically normal.

\paragraph{P2.\ Maximum Likelihood Estimator (MLE)}  Using the surrogate model

$$
  Y_i\mid Z_i\;\sim\;\text{Bernoulli}(\phi_i^{\!\top}\beta),
$$
take the parameter vector which maximizes the log-likelihood
$$
  \widehat\beta_{\text{MLE}}
  =\argmax_{\theta\in(0,1)^{|\mathcal R|}}
    \sum_{i=1}^{n}\Bigl[\,Y_i\log(\phi_i^{\!\top}\theta)
    +(1-Y_i)\log\bigl(1-\phi_i^{\!\top}\theta\bigr)\Bigr].
$$
Under the assumptions for OLS as well as an interior point assumption (INT) this estimator is consistent and asymptotically normal.

\paragraph{P3.\ \textsc{BIRDiE} Categorical--Dirichlet}
This estimator is a Bayesian treatment of the problem.  We keep the Bernoulli observation model under the true race
$$
  Y_i \mid Z_i, R_i \;\sim\; \operatorname{Bernoulli}\bigl(t_i^{\!\top}\beta\bigr),
$$
and give each race–specific approval probability a Beta prior
$$
\beta_r \sim \operatorname{Beta}(\alpha_1,\alpha_2),
\qquad r \in \mathcal{R}.
$$
This Beta-Bernoulli formulation is mathematically equivalent to \citeauthor{mccartan:2024}'s Categorical-Dirichlet model when the outcome is binary. We estimate the relevant posterior distribution over $\beta$ using $Y$ and $\Phi$. To derive a point estimate of $\beta$, we take the mode of the posterior distribution (the MAP estimate).
We include this estimator in the semi-synthetic study on HMDA data to illustrate equivalence in large-sample behavior between the OLS, MLE, and BIRDiE Categorical-Dirichlet methods.  BIRDiE provides additional flexibility for partial pooling over geographic areas through hierarchical models, which may improve estimation further when the (CI-YZ) assumption is violated by the geographic variable in particular. This is beyond the scope of this paper, but for further information on this method see \cite{mccartan:2024}.

\section*{Posterior Race Distributions}

Throughout, we consider two possibilities for deriving posterior race probabilities based on two distinct sets of  conditioning variables. Bayesian Improved Surname Geocoding estimates $\mathbb{P}(R = r \mid S_i, G_i)$ whereas Bayes Rule on Surname (BRS) estimates $\mathbb{P}(R = r \mid S_i)$. The BISG posterior uses both surname and tract under the additional assumption of conditional independence of surname and geography given race. The BRS posterior excludes geography to make the conditional independence assumption (CI–YZ) a weaker condition. (CI-YZ) is defined in the Assumptions section. The calculation for each posterior is as follows:

\[
\Phi^{\text{BISG}}_{i r}
  \;=\;
  \frac{\mathbb{P}\bigl(R = r\bigr)\,\mathbb{P}\bigl(S_i \mid R = r\bigr)\,\mathbb{P}\bigl(G_i \mid R = r\bigr)}
       {\displaystyle\sum_{r' \in \mathcal{R}} \mathbb{P}\bigl(R = r'\bigr)\,\mathbb{P}\bigl(S_i \mid R = r'\bigr)\,\mathbb{P}\bigl(G_i \mid R = r'\bigr)}
\quad\text{and}\quad
\Phi^{\text{BRS}}_{i r}
  \;=\;
  \frac{\mathbb{P}\bigl(R = r\bigr)\,\mathbb{P}\bigl(S_i \mid R = r\bigr)}
       {\displaystyle\sum_{r' \in \mathcal{R}} \mathbb{P}\bigl(R = r'\bigr)\,\mathbb{P}\bigl(S_i \mid R = r'\bigr)}.
\]

\section*{Data‑Generating Process}

The approval outcome is assumed to satisfy
\[
  Y_i \;\big|\, R_i \;\sim\;
  \operatorname{Bernoulli}\bigl(t_i^{\!\top}\beta\bigr),
  \qquad\text{that is}\qquad
  \mathbb{P}(Y_i=1 \mid R_i=r)=\beta_r .
\]
An individual’s approval probability equals the
race‑conditional approval probability for their true race.
\(Y_i\) is a finite mixture of Bernoulli
distributions—one component per race—with mixture weights
\(\mathbb{P}(R_i=r)\):
\[
  \mathbb{P}(Y_i=1)
  \;=\;
  \sum_{r\in\mathcal R}\mathbb{P}(R_i=r)\,\beta_r .
\]
Our target is the vector of race‑conditional approval
rates \(\beta\).

\medskip
\noindent\textbf{Surrogate model.}\;
Because the true race \(R_i\) is unobserved, we work with
a \emph{posterior race‑probability vector}
\(
  \boldsymbol{\phi}_i
  =( \Pr(R=r\mid Z_i) )_{r\in\mathcal R},
\)
where \(Z_i\) is an observed conditioning set
(e.g.\ \((S_i,G_i)\) for BISG or \(S_i\) for BRS). Thus we define the following \emph{surrogate model} 
\[
  Y_i \;\big|\, Z_i \;\sim\;
  \operatorname{Bernoulli}\!\bigl(\boldsymbol{\phi}_i^{\!\top}\beta\bigr),
  \qquad\text{equivalently}\qquad
  \mathbb{P}(Y_i=1 \mid Z_i)=\boldsymbol{\phi}_i^{\!\top}\beta .
\]
Replacing the unobserved one‑hot vector \(t_i\) by the observable
\(\boldsymbol{\phi}_i\) allows for the parameters of the mixture to
be estimated from observable data. In the following sections we demonstrate the conditions under which this likelihood is correctly specified and allows point identification of the true parameter $\beta$. The surrogate model implies a natural Maximum Likelihood Estimator \(\hat{\beta}_{\text{MLE}}\) as well as an Ordinary Least Squares estimator \(\hat{\beta}_{\text{OLS}}\), which can be used to estimate $\beta$ when the necessary assumptions hold.

\section*{Assumptions}

\begin{enumerate}[leftmargin=2.8em]

\item[(ACC)] Calibration of posterior race probabilities.
      For the conditioning set $Z_i$,
      \(
        \boldsymbol{\phi}_i
        :=\bigl(\mathbb{P}(R=r\mid Z_i)\bigr)_{r\in\mathcal R}
      \)
      satisfies
      \[
        \mathbb{E}\!\bigl[t_i \mid Z_i\bigr]=\boldsymbol{\phi}_i
      \] and assumptions of the race probability estimation method are satisfied.

      \begin{itemize}[leftmargin=1.8em]
        \item  BISG: $Z_i=(S_i,G_i)$ and 
               $S_i \;\perp\!\!\!\perp\; G_i \mid R_i$
        \item  BRS:  $Z_i=S_i$ with no further conditional independence assumptions.
      \end{itemize}

\item[(CI–YZ)] Conditional independence.\;
      \(Y_i \;\perp\!\!\!\perp\; Z_i \mid R_i\).

      \begin{itemize}[leftmargin=1.8em]
        \item  BISG: \(Y_i \;\perp\!\!\!\perp\; (S_i,G_i) \mid R_i\). CI-YZ applies jointly to surname and geography.
        \item  BRS:  \(Y_i \;\perp\!\!\!\perp\; S_i \mid R_i\). CI-YZ independence requirement is only with respect to surname, avoiding potential geographic violations.
      \end{itemize}

\item[(ID)] Full‑rank support.\;
      \(
        \Sigma_{\phi\phi}
          :=\mathbb{E}\!\bigl[\boldsymbol{\phi}_i
                              \boldsymbol{\phi}_i^{\!\top}\bigr]
      \)
      is nonsingular.

\item[(REG)] $\{(Y_i,Z_i)\}_{i=1}^n$ are sampled in an i.i.d. manner from their joint distribution.

\item[(INT)] Interior parameter.\;
      The true value $\beta$ lies in a compact subset of $(0,1)^{|\mathcal{R}|}$
      such that
      $0<\boldsymbol{\phi}_i^{\!\top}\theta<1$ almost surely\ for all $\theta$
      in a neighborhood of $\beta$.
\end{enumerate}

\medskip
\noindent
(ACC) and (CI–YZ) ensure correct specification, (ID) gives point
identification, while (REG) and (INT) provide the regularity
conditions used for the large‑sample results below.


\section*{Proof of Correct Specification and Identification}

\paragraph{Claim.}
Under (ACC), (CI-YZ), and (ID),  
the surrogate model  
\[
  Y_i \,\big|\, Z_i \;\sim\;
  \operatorname{Bernoulli}\!\bigl(\boldsymbol{\phi}_i^{\!\top}\beta\bigr)
\]
is correctly specified and the parameter vector
\(\beta=(\beta_r)_{r\in\mathcal R}\) is point‑identified.

\paragraph{Proof.}
By the law of total probability
\begin{align}
\mathbb{P}(Y_i=1 \mid Z_i)
  &=\sum_{r\in\mathcal R}
      \mathbb{P}(Y_i=1 \mid R_i=r,\,Z_i)\;
      \mathbb{P}(R_i=r \mid Z_i).
      \label{eq:totprob}
\end{align}

\medskip\noindent
\emph{Step 1 (conditional independence).}  
Assumption (CI-YZ) implies
\(\mathbb{P}(Y_i=1 \mid R_i=r,\,Z_i)
      =\mathbb{P}(Y_i=1 \mid R_i=r)=\beta_r\).
Insert this into \eqref{eq:totprob}:

\[
  \mathbb{P}(Y_i=1 \mid Z_i)
  =\sum_{r\in\mathcal R}\beta_r\,
     \mathbb{P}(R_i=r \mid Z_i).
\]

\medskip\noindent
\emph{Step 2 (calibration).}  
By (ACC),
\(\mathbb{P}(R_i=r \mid Z_i)=\phi_{i,r}\).
Therefore
\[
  \mathbb{P}(Y_i=1 \mid Z_i)
  =\sum_{r\in\mathcal R}\beta_r\,\phi_{i,r}
  =\boldsymbol{\phi}_i^{\!\top}\beta .
\]
This is exactly the Bernoulli mean in the surrogate model,
so the model is \emph{correctly specified}. In other words, the true parameter vector of the data generating process $\beta$ falls within the parameter space of the surrogate model.

\medskip\noindent
\emph{Step 3 (identification).}  
Assumption (ID) states that
\(\Sigma_{\phi\phi}=\mathbb{E}\bigl[\boldsymbol{\phi}_i
                              \boldsymbol{\phi}_i^{\!\top}\bigr]\)
is full rank.
Consequently, the linear function mapping
\(\beta\mapsto\Phi\beta\)
is injective, so \(\beta\) is point‑identified. Because of this, the maximum value of the objective function for the two types of extremum estimators proposed below (OLS and MLE) will be uniquely defined.
\hfill\(\square\)

\section*{Properties of Surrogate Model Estimators}

Throughout this section assume (ACC), (CI–YZ), (ID), (REG), and (INT)
hold.  The symbol \(\beta\) refers to the true population vector of
race‑conditional approval probabilities; \(\hat\beta\) denotes an
estimate computed from the sample.

\subsection*{OLS / Just‑Identified GMM}

We define the Ordinary Least Squares estimator in two equivalent ways per Hayashi \cite[p.~18]{hayashi:2000}.
We have the data-matrix for use in the finite-sample theory:
\[
  \hat{\beta}_{\mathrm{OLS}}
    \;=\;
    (\Phi^{\mathsf T}\Phi)^{-1}\Phi^{\mathsf T}Y.
\] 
For the large-sample, asymptotic theory we have the sample average form:
\[
  \hat\beta_{\text{OLS}}
    = S_{\phi \phi}^{-1}s_{\phi y} 
\]
where the two sample-averages are 
\[ 
    S_{\phi \phi} = \tfrac1n\sum_{i=1}^n
    \boldsymbol{\phi}_i\boldsymbol{\phi}_i^{\!\top}=\frac{\Phi^{\mathsf T}\Phi}{n}
\quad \text{and} \quad
    s_{\phi y} = \tfrac1n\sum_{i=1}^n
             \boldsymbol{\phi}_i Y_i = 
             \frac{\Phi^{\mathsf T}Y}{n}.
\]
The two forms are equivalent because the $\frac{1}{n}$ term is canceled by its multiplicative inverse $(\frac{1}{n})^{-1} = n$. The corresponding population moments for the sample average form are defined as
\[ 
    \Sigma_{\phi \phi} = \mathbb{E}[\boldsymbol{\phi}_i\boldsymbol{\phi}_i^{\!\top}]
\quad \text{and} \quad
    \Sigma_{\phi y} = \mathbb{E}[\boldsymbol{\phi}_i Y_i].
\]
Under our given assumptions the OLS estimator is unbiased, consistent, and asymptotically normal, which we will now demonstrate.

\paragraph{Finite-sample unbiasedness.}
Under the following assumptions, the Ordinary Least Squares (OLS) estimator is unbiased for every finite sample size \(n\) \cite[p.~27]{hayashi:2000}.

\begin{enumerate}[label=(OLS~\arabic*), leftmargin=2.8em]
  \item \textbf{Linearity.}  For each observation \(i = 1,\dots,n\) the relation between the regressors and the dependent variable is
  \[
    Y_i \;=\; \boldsymbol{\phi}_i^{\mathsf T}\beta + \varepsilon_i .
  \] 
  \emph{Justification}: Assume the linear model above. Under our surrogate model, $\varepsilon_i$ is derived as follows:
  \begin{align*}
    Y_i - \boldsymbol{\phi}_i^{\!\top}\beta &= \varepsilon_i \\
    Y_i - \mathbb{P}(Y_i=1 \mid Z_i) &= \varepsilon_i \quad \text{by proof of correct specification} \\
    Y_i - \mathbb{E}\bigl[Y_i \mid Z_i\bigr] &= \varepsilon_i \\
    Y_i - \mathbb{E}\bigl[Y_i \mid \Phi] &= \varepsilon_i \quad \text{by definition of the surrogate model} \\
  \end{align*}
  Thus, $\varepsilon_i$ is the difference between the realization of each Bernoulli random variable $Y_i$ and its expected value conditional on $Z_i$, or equivalently on $\Phi$.

  \item \textbf{Strict exogeneity.}  For each observation \(i = 1,\dots,n\) the mean of the error term conditional on the regressors is
  \[
    \mathbb{E}\bigl[\varepsilon_i \mid \Phi\bigr] \;=\; 0 ,
  \]

  \emph{Justification}:  We use our definition of $\varepsilon_i$ under the surrogate model.
  \begin{align*}
    \mathbb{E}\!\bigl[\varepsilon_i \mid \Phi\bigr]
      &= \mathbb{E}\!\Bigl[
           Y_i - \mathbb{E}\!\bigl[Y_i \mid \Phi\bigr]
           \,\Big|\, \Phi\Bigr] \\
      &= \mathbb{E}\!\bigl[Y_i \mid \Phi\bigr]
         \;-\;
         \mathbb{E}\bigl[\mathbb{E}\bigl[Y_i \mid \Phi\bigr] \mid \Phi\bigr] \quad \text{by linearity of expectation}\\
      &= \mathbb{E}\!\bigl[Y_i \mid \Phi\bigr]
         \;-\;
         \mathbb{E}\!\bigl[Y_i \mid \Phi\bigr] \quad \text{by idempotence of the conditional expectation}\\                   
      &= 0
  \end{align*}
  
  \item \textbf{No perfect multicollinearity.}
  The matrix \(\Phi\) has full column rank with probability 1. 
  This is true under the population full-rank support condition (ID) already assumed.
\end{enumerate}

\medskip\noindent
Thus,
\[
  \mathbb{E}\!\bigl[\hat{\beta}_{\mathrm{OLS}}\bigr] \;=\; \beta
\]
for any finite sample size \(n\).
\\\\
\emph{Note on other finite sample properties of OLS:}
Note that we do not assume the spherical error variance, which includes homoskedasticity and no correlation between residuals. While these assumptions lead to the Gauss-Markov theorem for the best linear unbiased estimator property and an expression for the variance of $\beta$, the residuals here are heteroskedastic. Because the outcome is Bernoulli, the variance of the error term $\varepsilon_i$ is fully determined by $\phi_i$ and $\beta$. 
Specifically, $\text{Var}(\varepsilon_i) = 
\text{Var}(Y_i - \mathbb{E}\bigl[Y_i \mid \Phi]) =
\mathrm{Var}(Y_i | \phi_i) =  p_i(1-p_i) = \mathbb{E}[Y_i | \phi_i]\cdot(1-\mathbb{E}[Y_i | \phi_i])$. 
For this reason, we will turn to a central limit theorem to describe the rate of convergence of $\hat{\beta}_{\text{OLS}}$ with respect to sample size.

\paragraph{Consistency and Asymptotic Normality.}
Under the following set of assumptions, \(\hat\beta_{\text{OLS}}\) is consistent and asymptotically normal per Hayashi \cite[pp.~113\,,\,209]{hayashi:2000}. We define the moment function
\[
  g_i(\beta) := \boldsymbol{\phi}_i (Y_i - \boldsymbol{\phi}_i^{\!\top} \beta),
\]
and verify the five assumptions required for consistency and asymptotic normality.

\begin{enumerate}[label=(OLS~\arabic*a), leftmargin=2.8em]
  \item \textbf{Linearity.} This has the same definition and justification as (OLS 1).
  \item \textbf{Ergodic Stationarity.} The sequence \(\{(Y_i, \boldsymbol{\phi}_i)\}_{i=1}^n\) is stationary and ergodic.

  \emph{Justification}: Under (REG), the sample consists of i.i.d. draws from the joint distribution of \((Y_i, \phi_i)\) which, per \citeauthor{hayashi:2000} \cite[p.~110]{hayashi:2000}, implies ergodic stationarity.

  \item \textbf{Orthogonality Condition.} For each observation, the regressors are orthogonal to the error term.
  \[
    \mathbb{E}[g_i(\beta)] = \mathbb{E}\big[\boldsymbol{\phi}_i (Y_i - \boldsymbol{\phi}_i^{\!\top} \beta)\big] = 0
  \]

  \emph{Justification}: Under the surrogate model we have
  \begin{align*}
    \mathbb{E}[g_i(\beta)] &= 
    \mathbb{E}\big[\phi_i (Y_i - \phi_i^{\!\top} \beta)\big] \\
    &= \mathbb{E}\big[\phi_i \varepsilon_i \big] \quad \text{by (OLS 1)} \\
    &= \mathbb{E}\big[\phi_i \mathbb{E}[\varepsilon_i \mid \phi_i] \big] \quad \text{by iterated expectations}\\
    &= \mathbb{E}\big[\phi_i 0 \big] \quad \text{by (OLS 2)} \\
    &= 0
  \end{align*}

  \item \textbf{Rank condition for identification.} This is the same as assumption (ID).

  \item \textbf{Martingale difference sequence with finite second moment.} The moment function \(g_i(\beta)\) is a martingale difference sequence with finite second moment and the covariance matrix \(\Sigma_{gg} := \mathbb{E}[g_i(\beta)g_i(\beta)^{\!\top}]\) is non-singular.
  
  \emph{Justification}: Under i.i.d. sampling per (REG) and orthogonality of the error term and regressors per (OLS 3a) the moment function is a martingale difference sequence. This is because under i.i.d. the data generating process has no serial dependence and orthogonality makes the expectation of the moment function zero. \[\mathbb{E}[g_i(\beta)| g_{i-1}(\beta),g_{i-2}(\beta),...,g_1(\beta)] = \mathbb{E}[g_i(\beta)] = 0\]
We demonstrate that the second moment matrix $\Sigma_{gg}$ has all finite values in the Appendix using an argument based its spectral norm.
Finally, nonsingularity of \(\mathbb{E}[\,g_i g_i^{\top}\,]\) follows from the
full–rank support condition (ID). We demonstrate this in the Appendix by proving that singular 
\(\mathbb{E}[\,g_i g_i^{\top}\,]\) implies singular \(\mathbb{E}[\,\phi_i \phi_i^{\top}\,]\), leading to a contradiction of assumption (ID).

\end{enumerate}
Given the assumptions above we have: 
\[
  \hat\beta_{\text{OLS}}\xrightarrow{p}\beta,
  \qquad
  \sqrt{n}\bigl(\hat\beta_{\text{OLS}}-\beta\bigr)
    \xrightarrow{d}
    N\!\Bigl(
       0,\;
       \Sigma_{\phi\phi}^{-1}\,
       \mathbb{E}[g_i(\beta) g_i(\beta)^T]
       \Sigma_{\phi\phi}^{-1}\Bigr).
\]
\emph{Note on Inference of Variance-Covariance Matrix:}
An extended discussion and analysis on  inference of the variance-covariance matrix of $\hat\beta_{\text{OLS}}$ is beyond the scope of this paper. We will note two requirements for the approach. First, it should be robust to the heteroskedasticity induced by this model's specification. This is addressable by \citeauthor{white:1980}'s heteroskedasticity consistent sandwich estimator, which is immediately deployable by \citeauthor{zeileis:2004}'s sandwich R package~\cite{white:1980}~\cite{zeileis:2004}. Second, it should consider whether treating posterior race probabilities as a known conditional probability rather than an estimate subject to sampling variability is acceptable. \citeauthor{lu:2024} addresses this sampling variability with their dual-bootstrap method and finds that the effect on variance estimation is very small in their simulations \cite{lu:2024}. Specifically, \citeauthor{lu:2024} conclude that ``measurement uncertainty is generally insignificant for BISG except in particular circumstances; bias, not variance, is likely the predominant source of error."

\subsection*{Maximum‑Likelihood Estimator}

The surrogate model is a generalized linear model with a Bernoulli distribution of the target variable and an identity link function.
\[
  Y_i \,\big|\, Z_i \;\sim\;
  \operatorname{Bernoulli}\!\bigl(\boldsymbol{\phi}_i^{\!\top}\beta\bigr)
\]
Recall that the parameter space is
\(
  \Theta=[0,1]^{|\mathcal R|}
\). The log-likelihood of $\theta \in \Theta$ is:
\begin{align*}
  \ell(\theta)
  &= \sum_i^n (Y_i \log\!\bigl(\phi_i^{\!\top}\theta\bigr) 
    + (1-Y_i) \log\!\bigl[1-\phi_i^{\!\top}\theta]) \quad \text{per i.i.d under (REG)}\\
  &= Y^{\!\top}\log\!\bigl(\Phi\theta\bigr)
     + (1-Y)^{\!\top}\log\!\bigl[1-\Phi\theta\bigr]
\end{align*}
A very natural estimator for $\beta$ under the surrogate model is simply the parameter vector $\hat{\beta}_\text{MLE}$ that maximizes the log-likelihood (and thus the likelihood) function.
\[
  \displaystyle
  \hat\beta_{\text{MLE}}
  =\arg\max_{\theta\in(0,1)^{|\mathcal{R}|}}\ell(\theta)
\]
The maximum of the log-likelihood function does not have a closed form. Despite this, the function is concave which allows for the application of efficient convex optimization under the interior parameter (INT) assumption. Suggested solvers for this optimization include Iteratively Reweighted Least Squares and Newton-Raphson \cite{green:1984} \cite{boyd:2004}. These solvers can use the known gradient and Hessian to more efficiently and accurately estimate the solution. We have the gradient
\[
  \nabla_\theta\ell(\theta) =
   \Phi^{\!\top}
   \Bigl(
      Y \,/\,(\Phi\theta)
      \;-\;
      (1-Y) \,/\,\bigl(1-\Phi\theta\bigr)
  \Bigr).
\]
and the Hessian
\[
  \nabla_\theta^{2}\ell(\theta)
  \;=\;
  -\,\Phi^{\!\top}\operatorname{diag}(Y \,/\,(\Phi\theta)^{\!2}
  \;+\;
  (1-Y) \,/\,\bigl(1-\Phi\theta\bigr)^{\!2})\,\Phi
\]
The MLE is consistent and asymptotically normal with asymptotic variance equal to the reciprocal of the Fisher Information under our stated assumptions.

\subsection*{Consistency and Asymptotic Normality of the MLE}

Under assumptions (MLE~1)–(MLE~10) per Proposition~7.3 (Consistency) and~7.8 (Asymptotic Normality) of Hayashi \cite{hayashi:2000},
\[
  \hat\beta_{\mathrm{MLE}}\xrightarrow{p}\beta,
  \qquad
  \sqrt{n}\bigl(\hat\beta_{\mathrm{MLE}}-\beta\bigr)
    \xrightarrow{d} N\!\bigl(0,\,I(\beta)^{-1}\bigr),
  \qquad
  I(\beta):=\mathbb{E}\!\bigl[-\nabla_\beta^{2}\ell_i(\beta)\bigr]
\]

\begin{enumerate}[label=(MLE~\arabic*), leftmargin=2.8em]

  \item \textbf{Ergodic stationarity.}  Same as (OLS 2a).

  \item \textbf{Compact parameter space.} $\Theta$ is a compact subset of $\mathbb{R}^{|\mathcal{R}|}$.
  
        \emph{Justification:} \(\Theta=[0,1]^{|\mathcal R|}\) is closed and bounded in
        \(\mathbb{R}^{|\mathcal R|}\) and is therefore compact by the Heine–Borel theorem.

  \item \textbf{Continuity of the log-likelihood in \(\beta\).} The log-likelihood is continuous in $\theta$ for each realisation.
  
        \emph{Justification:} For each realisation,  
        \(\beta\mapsto\ell_i(\beta)\)  
        is a composition of continuous functions
        evaluated away from the singularities
        0 and 1 by the interior-parameter assumption (INT);
        which implies continuity.

  \item \textbf{Measurability.} The log-likelihood is measurable in $(Y_i, \phi_i)$ for all $\theta \in \Theta$.
  
        \emph{Justification:} For all \(\theta 
        \in \Theta\) and each realisation \((Y_i,\phi_i)\) the function  
        \(\theta\mapsto\ell_i(\theta)\)  
        is continuous and therefore measurable.  

\item \textbf{Dominance}  
        For all \(\theta\in\Theta\) and for each realization \((Y_i,\phi_i)\), \(
\mathbb{E}\!\bigl[\lvert\ell_i(\theta)\rvert\bigr]<\infty
        \).
        \\\\
        \emph{Justification:}
Because \(Y_i\in\{0,1\}\) we can bound the absolute log-likelihood as  
\begin{align*}
\bigl|\ell_i(\theta)\bigr|
  &=\bigl|
      Y_i\log(\phi_i^{\!\top}\theta)
      +(1-Y_i)\log\bigl(1-\phi_i^{\!\top}\theta\bigr)
    \bigr| \\[2pt]
  &\le |Y_i|\:\bigl|\log(\phi_i^{\!\top}\theta)\bigr|
       +|1-Y_i|\:\bigl|\log\bigl(1-\phi_i^{\!\top}\theta\bigr)\bigr| \\[2pt]
  &\le \bigl|\log(\phi_i^{\!\top}\theta)\bigr|
       +\bigl|\log\bigl(1-\phi_i^{\!\top}\theta\bigr)\bigr| \\[2pt]
&\le \bigl|\log(|\epsilon|)\bigr| + \bigl|\log(1-|\epsilon|)\bigr| \quad \text{where \(\epsilon \in (0,1)\) per (INT)}\\[2pt]
 &< |\log(0)| = +\infty 
\end{align*}
Therefore
\[
\mathbb{E}\!\bigl[\,|\ell_i(\theta)|\,\bigr]
  <\;\infty
  \qquad \forall (\theta,Y_i, \phi_i) \in \{\Theta, Y, \Phi\}
\]
which verifies the required dominance condition.

\item \textbf{Interior true parameter.}  
      The true value \(\beta\) lies in the interior of \(\Theta\).
      \\\\
      \emph{Justification:}  
      This is assumption (INT).

\item \textbf{Twice differentiable log-likelihood.}  
      For every observation \((Y_i,\phi_i)\) the map
      \(\theta\mapsto\ell_i(\theta)\) is twice continuously differentiable
      on \(\Theta\).
      \\\\
      \emph{Justification:}  
      Follows from the derivation of the Hessian in the Appendix section “Derivation of Score and Hessian.”

\item \textbf{Score Central Limit Theorem.}  
      With
      \(s_i(\theta)=\nabla_\theta\ell_i(\theta)\) and
      \(\beta\) the true parameter,
      \[
        \frac{1}{\sqrt{n}}\sum_{i=1}^n s_i(\beta)
        \;\xrightarrow{d}\;
        N\!\bigl(0,\Sigma_s\bigr),
        \qquad
        \Sigma_s:=\operatorname{Var}\!\bigl[s_i(\beta)\bigr].
      \]
      \emph{Justification}  Per Hayashi \cite[p.~473]{hayashi:2000} a sufficient condition is that:
      (a) the data \(\{(Y_i,\phi_i)\}\) are i.i.d.\ (assumption REG), and  
      (b) the score has mean zero at the true parameter,
      \(\mathbb{E}[\,s_i(\beta)\,]=0\).
      \\
      Correct specification implies
      \(\mathbb{E}[Y_i\mid\phi_i]=\phi_i^{\!\top}\beta\); hence
      
\begin{align*}
\mathbb{E}\!\bigl[s_i(\beta)\mid\phi_i\bigr]
  &=\phi_i\;
    \mathbb{E}\!\Bigl[
       \tfrac{Y_i}{\phi_i^{\!\top}\beta}
       -\tfrac{1-Y_i}{1-\phi_i^{\!\top}\beta}
     \,\Big|\;\phi_i\Bigr] \\[4pt]
  &=\phi_i
     \Biggl[
       \frac{\mathbb{E}[Y_i\mid\phi_i]}{\phi_i^{\!\top}\beta}
       -\frac{\mathbb{E}[1-Y_i\mid\phi_i]}{1-\phi_i^{\!\top}\beta}
     \Biggr] \\[4pt]
  &=\phi_i
     \Biggl[
       \frac{\phi_i^{\!\top}\beta}{\phi_i^{\!\top}\beta}
       -\frac{1-\phi_i^{\!\top}\beta}{1-\phi_i^{\!\top}\beta}
     \Biggr]
     \qquad \text{by correct specification}
     \\[4pt]
  &=\phi_i(1-1)=0 .
\end{align*}

\item \textbf{Local dominance of the Hessian.}  
      There exists a neighbourhood \(N(\beta)\subset\Theta\) such that
      \[
        \sup_{\theta\in N(\beta)}
        \mathbb{E}\!\bigl[
          \lVert\nabla_\theta^{2}\ell_i(\theta)\rVert
        \bigr]\;<\;\infty .
      \]
      \emph{Justification.}  
      The per-observation Hessian is
      \[
        \nabla_{\theta}^{2}\ell_i(\theta)
        =-\phi_i\phi_i^{\!\top}\;
          d_i(\theta),\qquad
          d_i(\theta)=
          \frac{Y_i}{(\phi_i^{\!\top}\theta)^{2}}
          +\frac{1-Y_i}{(1-\phi_i^{\!\top}\theta)^{2}} .
      \]

      Assumption (INT) provides an \(\varepsilon\in(0,1)\) such that  
      \(\varepsilon\le\phi_i^{\!\top}\beta\le1-\varepsilon\) almost surely.  
      Define the axis-aligned hyper-cube
      \[
        N(\beta)
        :=\Bigl\{
              \theta\in\Theta:
              \lVert\theta-\beta\rVert_\infty\le\varepsilon/2
           \Bigr\}.
      \]
      For any \(\theta\in N(\beta)\) we have  
      \(\bigl|\phi_i^{\!\top}\theta-\phi_i^{\!\top}\beta\bigr|
       \le\lVert\theta-\beta\rVert_\infty\le\varepsilon/2\),
      and therefore
      \[
        \varepsilon/2\;\le\;\phi_i^{\!\top}\theta\;\le\;1-\varepsilon/2
        \quad\text{a.s.}
      \]
      Because \(Y_i\in\{0,1\}\), this implies
      \[
        0<d_i(\theta)\le
        \frac{1}{(\varepsilon/2)^{2}}
        +\frac{1}{(1-\varepsilon/2)^{2}}
        =:D_{\max}<\infty .
      \]
      Meanwhile \(\lVert\phi_i\phi_i^{\!\top}\rVert=\lVert\phi_i\rVert^{2}\le1\)
      since the components of \(\phi_i\) are non-negative and sum to 1.
      Hence, for every \(\theta\in N(\beta)\),
      \[
        \lVert\nabla_{\theta}^{2}\ell_i(\theta)\rVert
        \le
        \lVert\phi_i\phi_i^{\!\top}\rVert\,d_i(\theta)
        \le D_{\max}.
      \]
      Taking expectations and a supremum over \(\theta\in N(\beta)\) gives  
      \[
        \sup_{\theta\in N(\beta)}
        \mathbb{E}\!\bigl[
          \lVert\nabla_\theta^{2}\ell_i(\theta)\rVert
        \bigr]
        \le D_{\max}<\infty,
      \]
      establishing local dominance.

\item \textbf{Non-singular information matrix.}  
      The Fisher information at the true parameter,
      \[
        I(\beta)
        :=\mathbb{E}\!\bigl[-\nabla_{\beta}^{2}\ell_i(\beta)\bigr],
      \]
      is nonsingular.

      \medskip\noindent
      \emph{Justification:}  In the appendix we give a proof by contradiction similar to the non-singularity proof for the covariance matrix of the moment condition. Assuming a singular information matrix at the true parameter contradicts (ID), therefore the matrix is non-singular.
\end{enumerate}
\emph{Note on Inference of Variance-Covariance Matrix:}
While inference of the variance-covariance matrix of $\hat{\beta}_{\text{MLE}}$ is beyond the scope of this paper, as we suggest that the approach taken might benefit from robustness to small violations of the assumptions for correct specification as discussed in \cite{white:1982}. As with OLS, the uncertainty in posterior race probabilities should be considered when selecting and validating a variance-covariance matrix estimator.

\section*{Simulation Study}

In this simulation study we examine the convergence behavior of the heuristic and principled estimators under three scenarios that conform to the assumptions of the principled methods and two scenarios that violate them. The true label, localized probability, and uniform probability scenarios check convergence under different levels of entropy in the conditional distributions $\mathbb{P}\bigl(R_i \mid S_i\bigr)$ and $\mathbb{P}\bigl(R_i \mid G_i\bigr)$ given that the assumptions hold. The (CI-YZ) and (ID) violation scenarios each examine sensitivity of the estimate to violations of the assumption. For the time being, the (ID) violation is only discussed conceptually.

\subsection*{Experiment Overview}

In each scenario we repeatedly generate a true race matrix $T$, a random BISG posterior race probability matrix $\Phi^{\text{BISG}}$, and a loan approval vector $Y$ for a sample of $n$ applications. We then calculate $\hat{\beta}$ under each of the estimators compared. This constitutes a single Monte Carlo iteration. Details for the data generation algorithm under each scenario are given in \Cref{app:sim}.

For each Monte Carlo iteration, total variation distance (TVD) between true $\beta$ and $\hat{\beta}$ is calculated for each estimator. For the (CI-YZ) violation scenario we use the sample mean conditional on true race under $n=50,000$ to closely approximate true $\beta$ since $\beta$ is not fixed. In the true label, localized probability, and uniform probability cases we repeat this across selected sample sizes between $n=100$ and $n=50,000$, resulting in an empirical curve illustrating the convergence behavior of each estimator. We repeat this for 1,000 iterations (using a random seed for reproducibility) and report the average value of the TVD for each estimator for the given sample size. 

We apply the simplifying assumptions that there are 5 races, 5 geographic areas, and 5 surnames. Prior race probability is set to be equal across all races; \(\forall r \in \mathcal{R}. \; \mathbb{P}(R=r)=1/5\ \). In practice surname and geography are higher dimensional and prior race probability estimation must be carefully handled (as we see in the next section on Los Angeles Home Mortgage Approvals). We fix the true race conditional approval rates for all but the (CI-YZ) violation case to be $\beta = (0.80, 0.50, 0.60, 0.20, 0.55)$. 

The estimators compared include all of the heuristic and principled methods introduced in "Heuristic and Principled Estimators" with the exception of BIRDiE. All estimates for this experiment are made under the BISG form of the posterior probability and the simulated data generating process is consistent with BISG's assumptions. For the Threshold method, a $\tau$ value of 0.5 is chosen. The methods corresponding to their names in the plot legends for this section are as follows:

\begin{enumerate}[label=\arabic*.]
\item Maximum A‐Posteriori under BISG (\textit{MAP})
\item $0.5$ Threshold under BISG (\textit{Threshold})
\item Weighting Estimator under BISG (\textit{Weighting})
\item Surrogate OLS under BISG (\textit{OLS})
\item Surrogate MLE under BISG (\textit{MLE})
\end{enumerate}

\subsection*{Total Variation Distance}

Consider a probability measure in which an event is a set of tuples combining race and approval/decline. The sample space has 10 elementary outcomes, one for each race–decision pair. An event in this space is a set of those outcomes such as: $$\{(\text{White}, 0), (\text{White}, 1), (\text{Hispanic}, 0), (\text{Hispanic}, 1)\}.$$ The race-conditional approval rates $\mathbb{P}(Y=1 \mid R=r)$ and the race probabilities $\mathbb{P}(R=r) = 1/5$ are sufficient to assign probabilities to all events in the space. Since race probability is a constant, we denote the measure as $M(u)$ for a given vector $u = (u_1, ..., u_5)$ where $u_r = \mathbb{P}(Y=1 \mid R=r)$. The total variation distance between two such measures, $\text{TVD}( M(u), M(v) )$ is defined as the maximum possible difference in the probability that they assign to the same event. Following McCartan et al \cite{mccartan:2024}, we choose this as a convenient one-dimensional measurement of the difference between two such probability measures. In our special case the total variation distance between two measures simplifies to the average absolute difference between the two conditional approval probability vectors. The derivation for this fact is provided in \Cref{app:proof-tvd}.
\[
  {\text{TVD}}(M(u),M(v))
  \;=\;
  \frac{1}{5} \sum_{r \in \mathcal{R}} \bigl|u_r - v_r\bigr|
\]

\subsection*{The \emph{True Label} Scenario}

\begin{figure}[H]
  \centering
  \includegraphics[width=0.8\linewidth]{"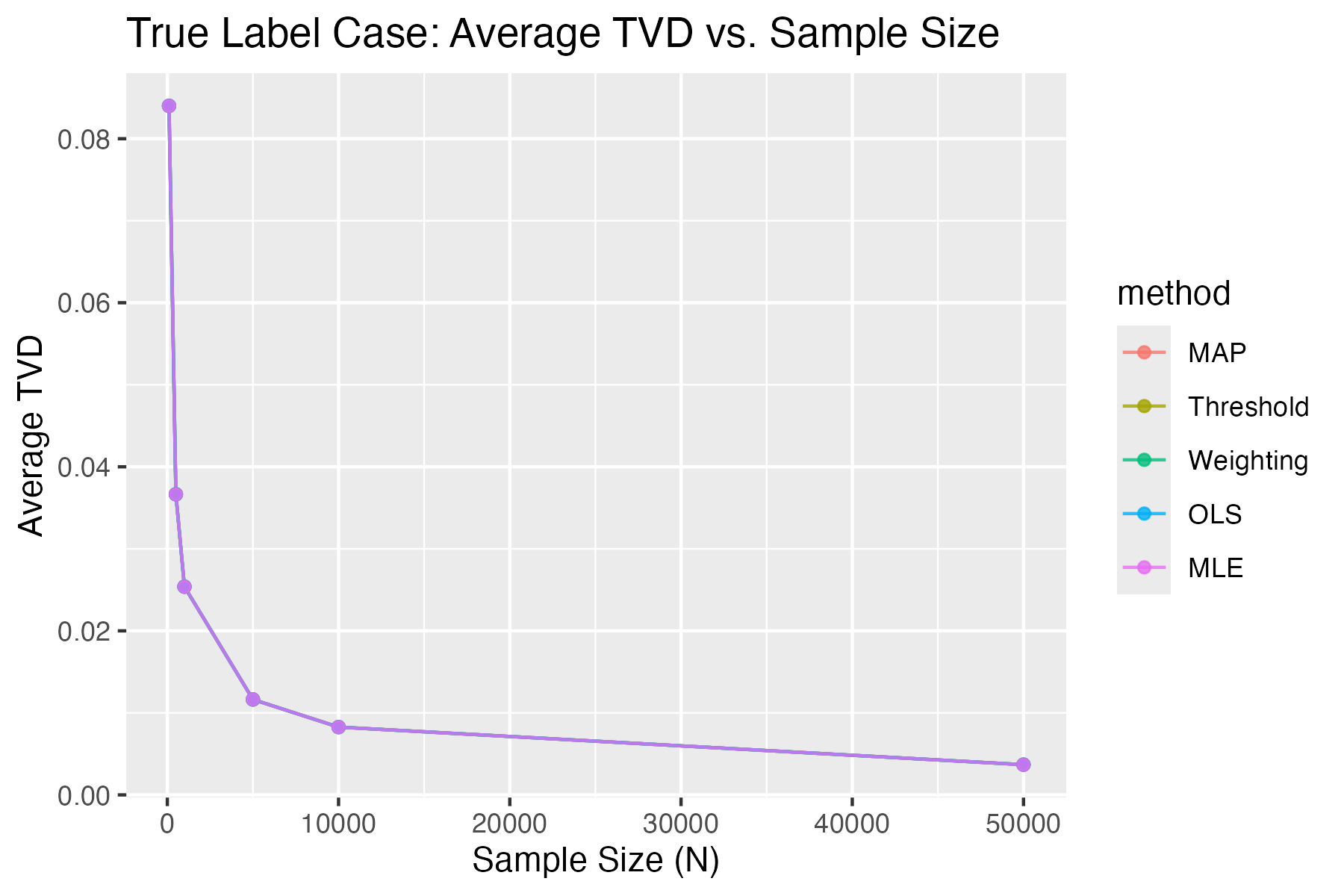"}
  \caption{Error under True Labels}
  \label{fig:true_label}
\end{figure}

In the true label scenario, each probability of race is either 1 or 0. This would occur, for instance, in mortgage lending where the self-reported race can be treated as true and known. In this case, all of the estimators behave exactly the same as the race-conditional sample mean. For the Threshold and MAP methods, predicted race from the classification rule is equal to true race. The Weighting estimator, being a generalization of the conditional sample mean, becomes the conditional sample mean in this case. Since the sample mean is the least squares estimator, the OLS coefficients are identical to this. Since the conditional sample mean is the maximum likelihood estimate for the race-conditional approval probability, the MLE estimates are also identical to this. This is illustrated by the identical curves in Figure \ref{fig:true_label}.

\subsection*{The \emph{Localized Probability} Scenario}

\begin{figure}[H]
  \centering
  \includegraphics[width=0.8\linewidth]{"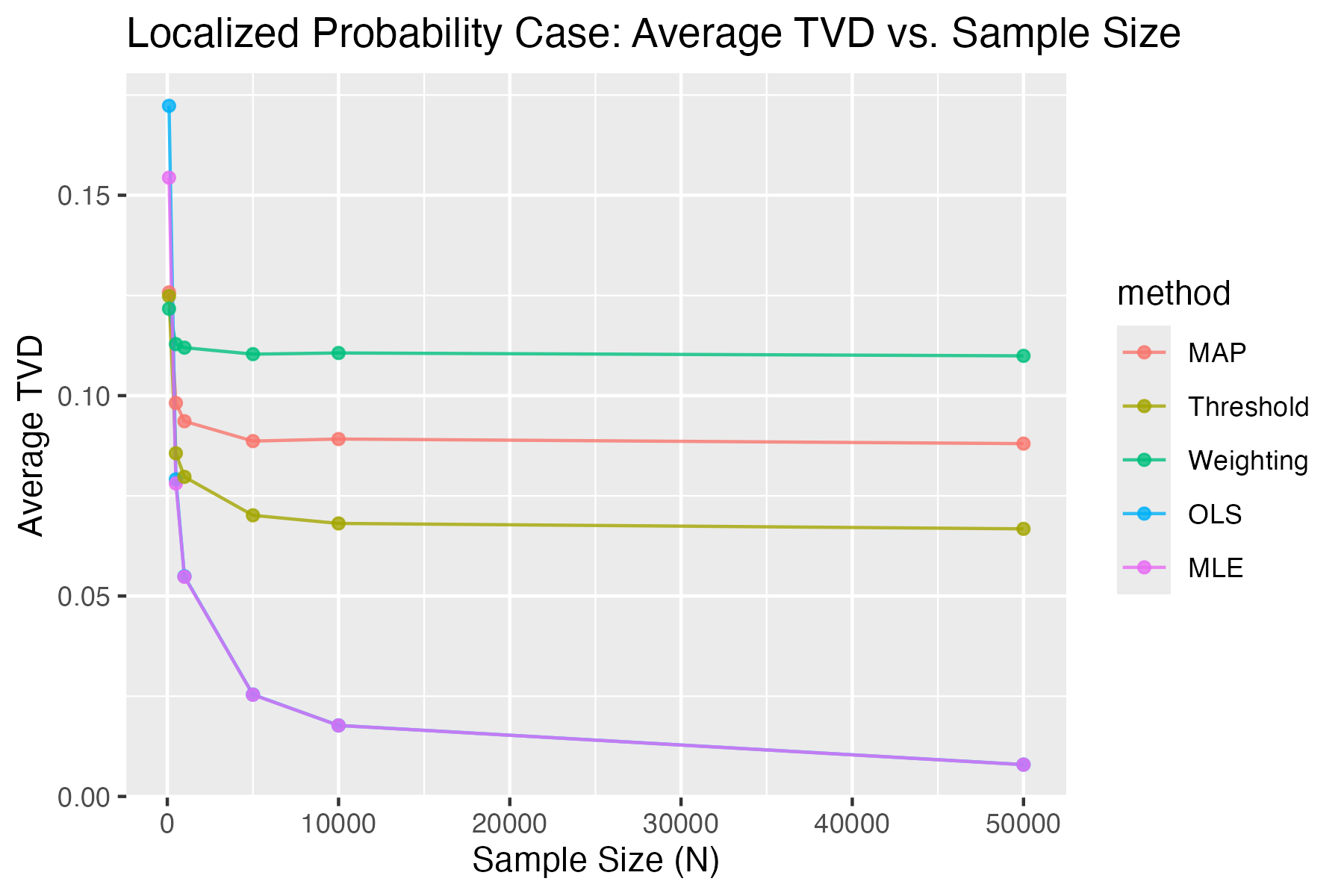"}
  \caption{Error under Localized Probabilities}
  \label{fig:localized}
\end{figure}

We consider a case similar to real-world BISG probabilities where the entries of the matrix $\Phi$ are random, but still localized. In this context, we mean by localization that for each row, it is fairly likely for one entry of the group probabilities to be significantly larger than the others. A detailed description of the simulated data generation process is provided in \cref{app:sim}. Intuitively, this situation corresponds to the case where the information about a person
belonging to a certain group is relatively
reliable. Figure~\ref{fig:localized} shows the results. As expected, the principled estimators show convergence of the TVD toward zero. The heuristic estimators converge toward a parameter that is other than the target parameter, even if we increase the sample size arbitrarily.

\subsection*{The \emph{Uniform Probability} Scenario}

\begin{figure}[H]
  \centering
  \includegraphics[width=0.8\linewidth]{"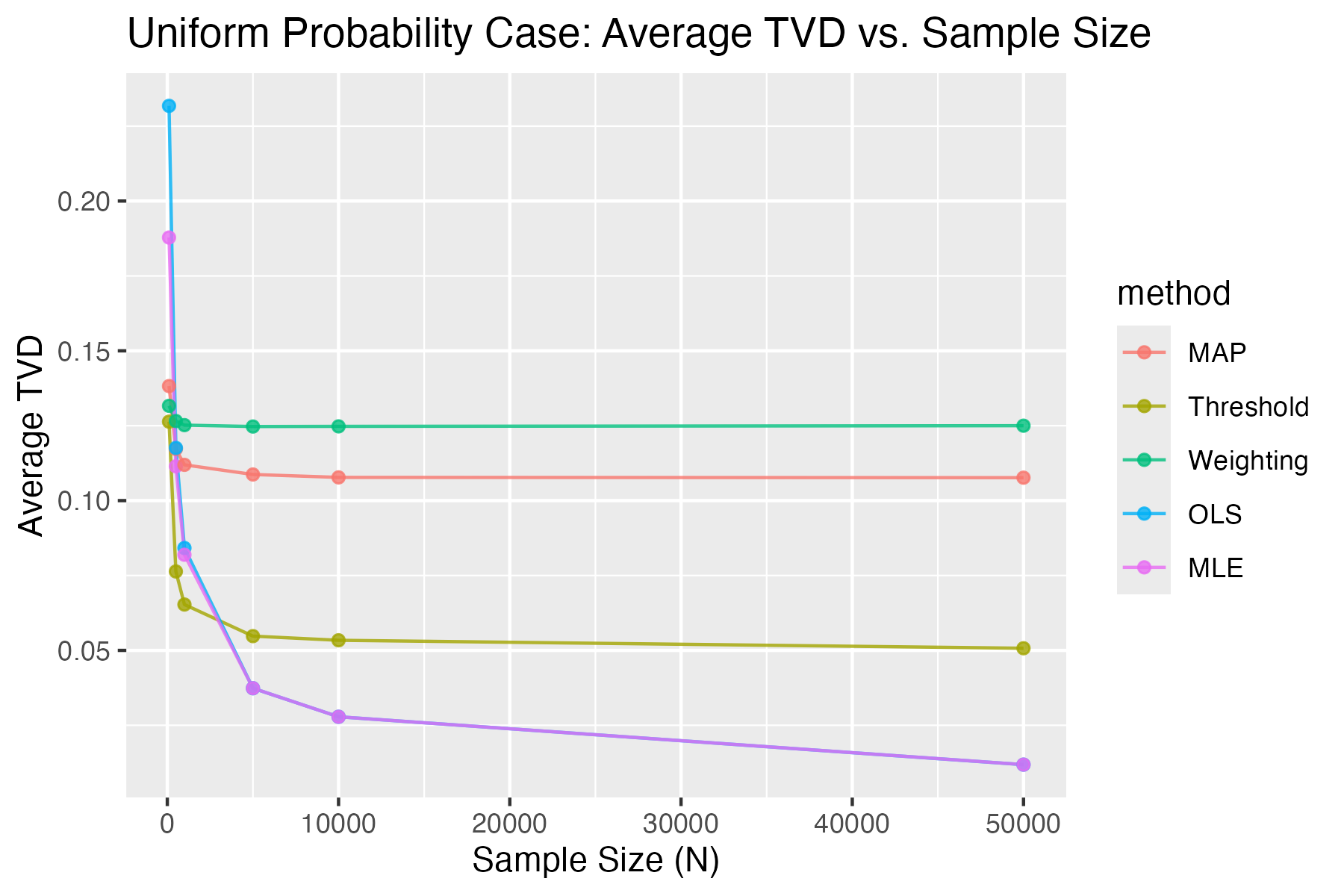"}
  \caption{Error under Uniform Probabilites}
  \label{fig:uniform}
\end{figure}

Imagine now a case where the rows of the matrix $\Phi$ are points sampled with uniform probability from the probability simplex for 5 categories. This space is a 4-simplex in $\mathbb{R}^5$ covering the set of all possible race probability values for a given individual. The probabilities in this case are not localized, and as such, we expect slower convergence amongst principled estimators and a different asymptotic bias amongst the heuristic estimators. The results are presented in \Cref{fig:uniform}. 

\subsection*{The \emph{(CI-YZ) violation} scenario} 

\begin{figure}[H]
  \centering
  \includegraphics[width=0.8\linewidth]{"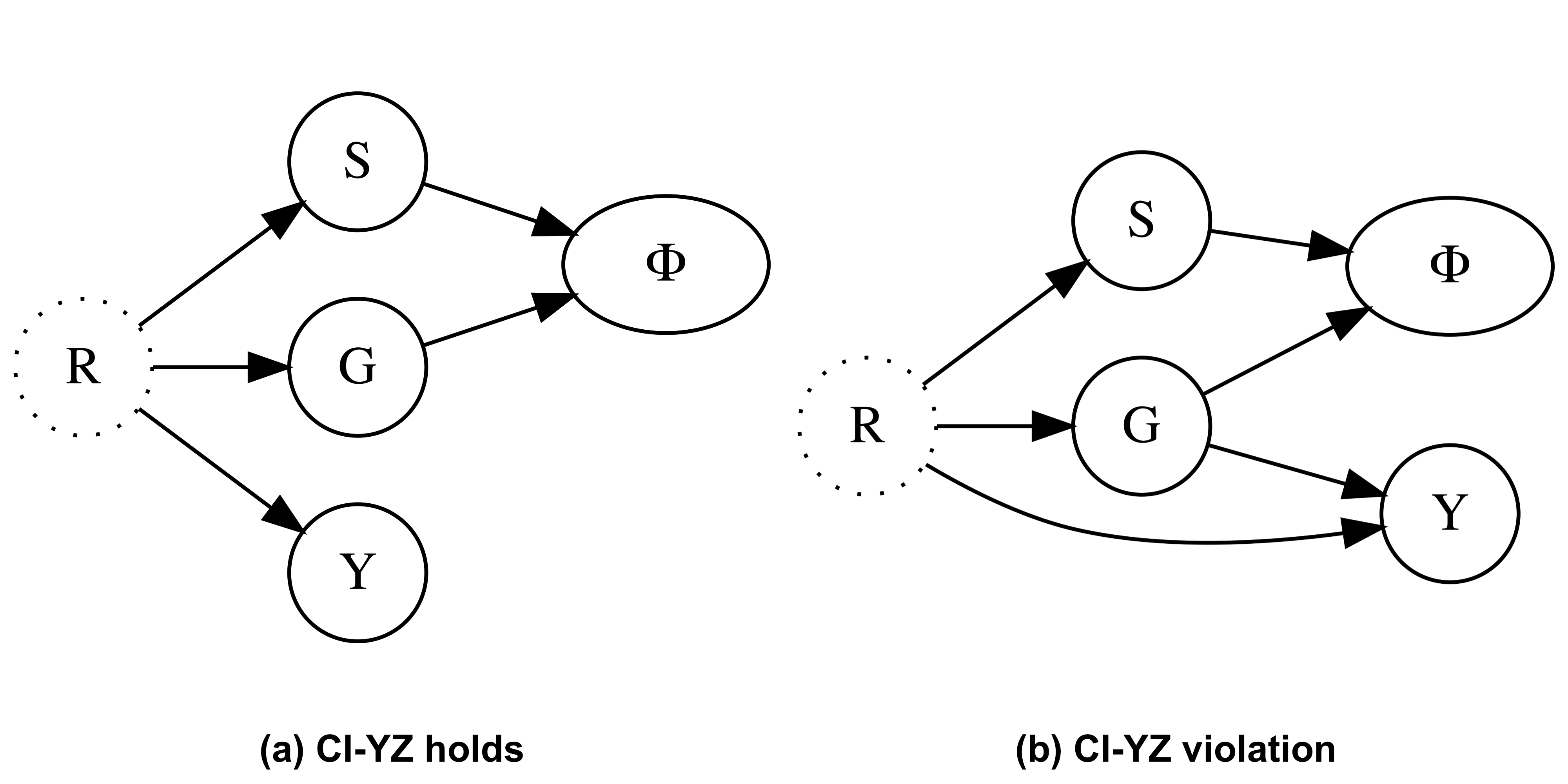"}
  \caption{Causal Diagrams for CI-YZ}
  \label{fig:causal_dag}
\end{figure}

Here we assume the localized probability scenario but alter the data generating process such that we violate assumption (CI-YZ). We accomplish this by adding a geographic effect to the loan approval probability that is not mediated by race. Figure \ref{fig:causal_dag} presents two contrasting causal directed acyclic graphs. In \ref{fig:causal_dag}a surname and geography are proxies for race but do not affect approval probabilities directly. In \ref{fig:causal_dag}b geography has a direct effect on approval probability as well as being a proxy for race. If \ref{fig:causal_dag}b is the true diagram, the OLS and MLE methods will fail to identify the true parameter and an asymptotic bias will be introduced to the estimate. 

To study what happens under such a violation of the assumption, we question how the total variation distance will respond to an additive geographic effect $\alpha$. We choose the sample size $n=50,000$ to approximate asymptotic behavior. We define $\alpha$ as the extent to which being in Geography 5 (arbitrarily chosen) increases (or decreases) the probability of approval. We define $\gamma$ as the race-conditional approval rates prior to the adjustment and fix $\gamma = (0.80, 0.50, 0.60, 0.20, 0.55)$. We choose values of $\alpha \in (-0.55,0.45)$ such that $0 < \mathbb{P}(Y_i=1 \mid R_i = 5) < 1$.

\[
\mathbb{P}(Y_i = 1)
  \;=\;
  t_i^{\!\top}\gamma
  + 
  \alpha\,\mathbf{1}\{G_i = 5\},
\]

\begin{figure}[H]
  \centering
  \includegraphics[width=0.8\linewidth]{"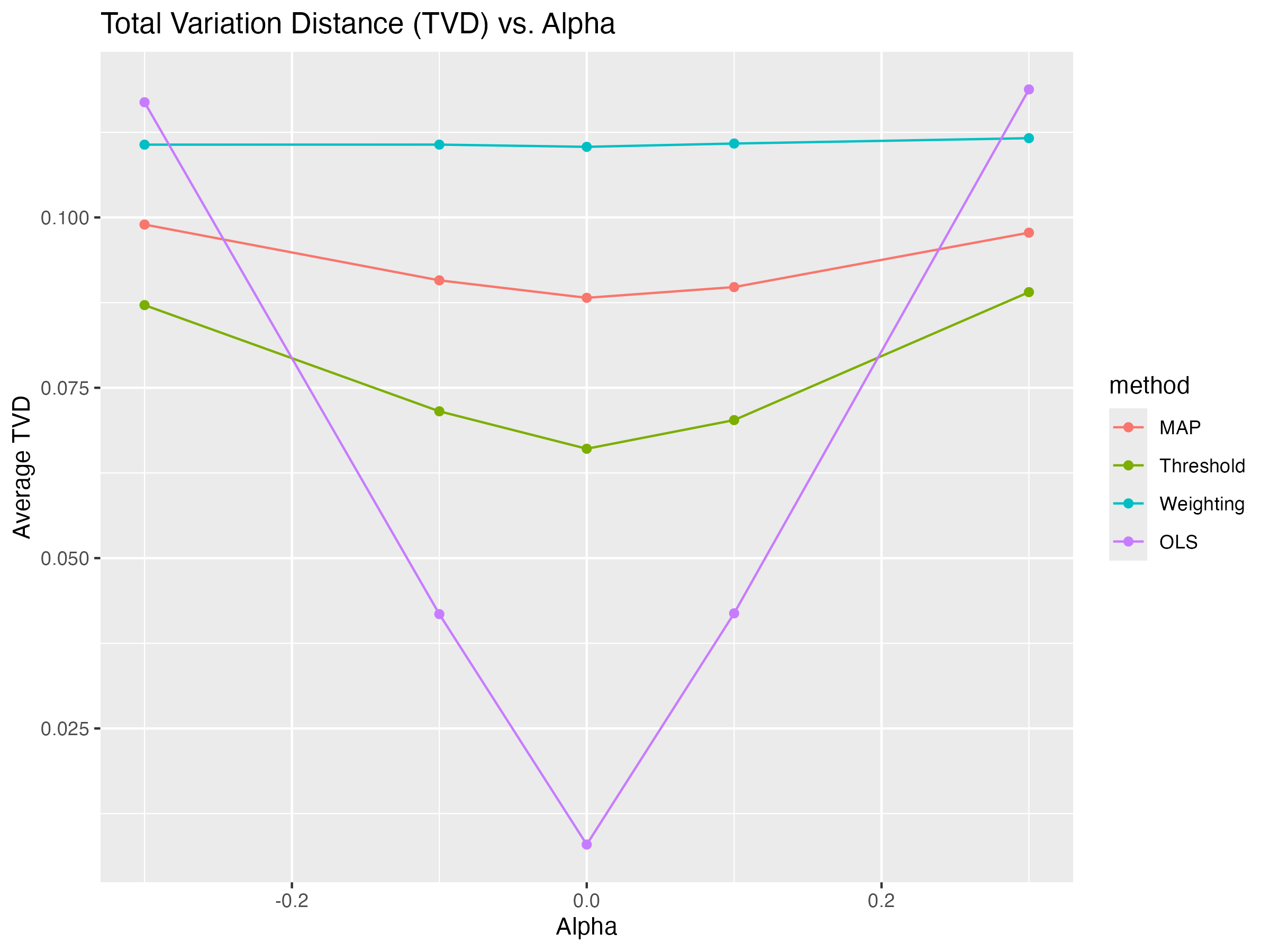"}
  \caption{Total Variation Distance w.r.t. $\alpha$}
  \label{fig:violation}
\end{figure}

$\beta$, the race-conditional approval rate after adjustment, then depends on the $\mathbb{P}(G_i = 5 | R_i)$. Under the localized random matrix generation process, $\mathbb{P}(G_i=5|R_i)$ has most of its probability concentrated in Race 5 with the distribution amongst the remaining Races being uniform. We use the race-conditional sample average under the true race as an approximation to the true $\hat{\beta}^{\text{true}} \approx \beta$, and measure the total variation distance relative to this sample average, $\text{TVD}(\hat{\beta}^{\text{method}}, \hat{\beta}^{\text{true}})$. \Cref{fig:violation} illustrates the total variation distance at several values of $\alpha$. As we can see, the OLS is more sensitive than the heuristics to $\alpha$, and, in this case, a geographic distortion on the order 30pp in approval rate is able to increase their averaged TVD beyond that of the other methods. As the OLS and MLE are both consistent, we assume they are similarly sensitive to $\alpha$ at $n=50,000$.

\subsection*{The \emph{(ID) violation} scenario}

In all cases we have so far considered, the matrix $\Phi^{\text{BISG}}$ was
non-singular, assumption (ID), which allowed the principled methods to accurately
estimate the group-specific acceptance rates given sufficient sample size.
However, we may violate the assumption (ID) by making the probability matrix numerically near-singular, in which case it will be impossible to determine the race-conditional acceptance rates in practice. 

Imagine an extreme scenario in which the posterior race probability matrix has equal entries. Then, clearly, no information regarding the race-conditional acceptance rates can be extracted from the data. What about cases that are not quite that extreme, but are close? As the $\mathbb{P}(R_i \mid S_i, G_i)$ tend toward maximum entropy, fully uniform, and thus uninformative probabilities it becomes less and less possible to attribute a given approval outcome $Y_i$ amongst the races $r \in \mathcal{R}$. How can we measure the suitability of the posterior race probabilities for disparity estimation and avoid this pitfall?

The condition number $\kappa(\Phi)$ provides an important measure for how reliable the computation of the race-specific acceptance rates is expected to be. Specifically, it tells us the sensitivity of $\hat{\beta}$ with respect to small changes in $Y$. A high condition number implies high volatility of $\hat{\beta}$ under repeated draws of the data from the data generating process. In a future experiment we plan to examine the relationship between the condition number, the variance of $\hat{\beta}$, and the estimation error in this setting. We recommend that analysts use the condition number to safeguard against proceeding with unreliable point estimation regardless of the choice of estimator.

\section*{Semi-synthetic Study: Los Angeles Home Mortgage Approvals}

The loan-level mortgage approval, geography (census tract), and race/ethnicity data made available by the FFIEC under the  Home Mortgage Disclosure Act offer a publicly available dataset suitable for evaluating the accuracy of these race-conditional approval rate estimators. However, because the surname variable required for BISG is not available in the data, we introduce a synthetic surname variable by Monte Carlo sampling from $\mathbb{P}(S=s\mid R=r)$ such that it satisfies the conditional independence assumption (CI-YZ). We then analyze the distribution of estimates over 10,000 surname imputation Monte Carlo iterations. 
Whilst $\beta$ is not strictly known, the sample averages of approval grouped within true race categories give us an unbiased and efficient estimate that surpasses the efficiency attainable by any method relying on imperfect posterior race probabilities. We define $\betatrue$ as follows:
\[
  \betatrue
  = \bigl(T^{\top}T\bigr)^{-1} T^{\!\top} Y
  = D^{-1} T^{\top} Y
  \quad \text{where} \quad
  D := \operatorname{diag}(n_1,\dots,n_{|\mathcal R|}).
\]
We use $\betatrue$ as a substitute for $\beta$, assuming that the large sample sizes (see next section) allow for the variance of $\betatrue$ to be a negligible source of error. In addition to the performance of each estimation methodology, we explore how the post-stratification of the $\mathbb{P}(R)$ prior by income can improve disparity estimation accuracy.

\subsection*{Data Overview}

We assess each estimator on the
\textit{2023 Home Mortgage Disclosure Act (HMDA)} public release for the
Los Angeles–Long Beach–Anaheim metropolitan statistical area, which is Core Based Statistical Area (CBSA) 31080 \cite{ffiec:hmda2023}.
We define the population considered to be first-lien, conventional, site-built,
1–4 unit purchase-mortgage applications. We harmonize HMDA categories
with those of BISG per the following table:
\[
\begin{array}{ll}
\text{HMDA category} & \text{BISG label} \\ \hline
\text{Hispanic or Latino (ethnicity)}                       & \text{hispanic} \\[4pt]
\text{White}                                                & \text{white}    \\[4pt]
\text{Black or African American}                            & \text{black}    \\[4pt]
\text{Asian \text{ or } Native Hawaiian or Other Pacific Islander} & \text{api} \\[4pt]
\text{American Indian or Alaska Native,\; 2{+} minority races,\; Joint,\; Free Form Text Only} & \text{other}
\end{array}
\]
Other is the union of the American Indian and Alaska Native (AIAN) group as well as the multi-racial group for BISG. This choice is made because the AIAN group had only 108 applicants, making sufficiently accurate approval rate estimation infeasible. The resulting sample contains $54\,248$ total applications with binary approval outcomes
$Y_i\in\{0,1\}$ and five racial groups with sample sizes $n_{\text{white}} = 23{,}008,\;
 n_{\text{black}} = 1{,}823,\;
 n_{\text{API}}   = 17{,}031,\;
 n_{\text{hispanic}} = 9{,}432,\;
 n_{\text{other}} = 2{,}954$.

\paragraph{Conditional race frequencies.}
Posterior surname–race probabilities
$\mathbb{P}(R=r\mid S=s)$ are taken from
the 2010 Census surname file given in the CFPB code repository for their BISG implementation \cite{census:2010_surnames}. We then calculate the following for use in generating surnames given true race as well as BISG/BRS.
\[
\mathbb{P}(S = s \mid R = r) \;=\;
\frac{\mathbb{P}(R = r \mid S = s)\,\mathbb{P}(S = s)}
     {\displaystyle\sum_{s' \in \mathcal{S}}
        \mathbb{P}(R = r \mid S = s')\,\mathbb{P}(S = s')}
\]
Race proportions at the census-tract level
$\mathbb{P}(R=r\mid G=g)$ are computed from the
2023 ACS 5-year table B03002 at the tract level \cite{census:acs5_b03002_2023}. The groups Asian and NHPI are combined into BISG category "API" at this step. The same formula above is applied to derive $\mathbb{P}(G=g\mid R=r)$.

\paragraph{Three alternatives for prior race frequencies.}
We compare three specifications for the marginal prior
$\pi_r=\mathbb{P}(R=r)$. Pace, in "The Hidden Biases in BISG Proxy-based Disparity Estimates" notes that "mortgage applicants tend to be of higher economic quality than the general population of U.S. adults
on which the BISG proxy model is based" \cite{pace:2021}. This poses a significant violation of the (ACC) assumption under standard BISG. We mitigate but do not solve this violation through the use of an income-stratified prior race distribution intended to better match the economic characteristics of the mortgage applicant population. 

\begin{enumerate}[label=\Alph*.]
\item \textbf{Standard} — the 2023 ACS CBSA-wide race mix taken from table b03002 at the MSA level \cite{census:acs5_b03002_2023}
\item \textbf{Income-Stratified} — A weighted average of the 2023 ACS race mix \emph{within}
      borrower-income bands, with weights estimated to match the proportion of loan applications in each band.
      $$\mathbb{P}(R=r) = \sum_{b} w_b\,\mathbb{P}(R=r \mid b) \quad \text{where b represents each income band}$$
      The $\mathbb{P}(R=r \mid b)$ frequency table is derived from ACS 5 five year table b19001 \cite{census:acs5_b19001_2023}.
      The weights are estimated as the proportion of the sample in each band in the HMDA data sample and are given below.
      \begin{table}[ht]
        \centering
        \begin{tabular}{lrr}
        \textbf{Income Band} & \textbf{$n_b$} & \textbf{$w_b$}\\
        $\le{}75$k   & 5\,633 & 10.4\%\\
        75--100k     & 2\,706 & 5.0\%\\
        100--125k    & 4\,353 & 8.0\%\\
        125--150k    & 4\,846 & 8.9\%\\
        150--200k    & 9\,289 & 17.1\%\\
        200k{+}      & 27\,421 & 50.6\%\\
        \end{tabular}
        \caption{HMDA borrower income distribution and corresponding weights.}
        \label{tbl:bands}
        \end{table}

\item \textbf{Perfect (Artificial)} — the empirical race mix of the
      HMDA sample. We treat this as an oracle to understand what would happen under a prior that exactly satisfied (ACC).
\end{enumerate}

The resulting priors are
\begin{table}[ht]
\centering
\begin{tabular}{lrrrrr}
\toprule
          & $\mathbb{P}(\text{api})$
          & $\mathbb{P}(\text{black})$
          & $\mathbb{P}(\text{hispanic})$
          & $\mathbb{P}(\text{other})$
          & $\mathbb{P}(\text{white})$ \\
\midrule
Standard            & 16.8\% & 6.1\% & 45.1\% & 3.7\% & 28.3\% \\
Income-Stratified   & 15.9\% & 5.0\% & 26.0\% & 11.5\% & 41.6\% \\
Perfect             & 31.4\% & 3.4\% & 17.4\% & 5.5\% & 42.4\% \\
\bottomrule
\end{tabular}
\caption{Race-probability priors and the observed applicant distribution (percentages).}
\end{table}

\paragraph{Estimators Compared.}
Eight estimators are evaluated combining the methods introduced in "Heuristic and Principled Estimators" with BISG and/or BRS. BRS is applied only to the OLS and MLE methods. The BIRDiE Categorical Dirichlet model is specifically used in conjunction with BISG by its definition \cite{mccartan:2024}. For the Threshold method, a $\tau$ value of 0.5 is chosen. 

\begin{enumerate}[label=\arabic*.]
\item Weighting Estimator (\textit{Weighting, BISG})
\item Maximum A‐Posteriori (\textit{MAP, BISG})
\item $0.5$ Threshold (\textit{Threshold, BISG})
\item Surrogate MLE (\textit{MLE, BISG})
\item Surrogate OLS (\textit{OLS, BISG})
\item \textsc{BIRDiE} Categorical Dirichlet, completely pooled model (\textit{BIRDiE, Cat-Dir})
\item Surname-only Surrogate MLE (\textit{MLE, BRS})
\item Surname-only Surrogate OLS (\textit{OLS, BRS})
\end{enumerate}

\paragraph{Surname Monte Carlo Procedure.}

Given the fixed outcomes $Y_i$, true race $R_i$ , and geography $G_i$, we run
$K=10{,}000$ Monte Carlo iterations. In each iteration we generate a synthetic surname
for each record by drawing once from $\mathbb{P}(S_i = s \mid R_i=r)$ for each record. We 
then measure the following for each selected combination of estimator and posterior probability matrix.

\begin{itemize}[noitemsep]
\item estimated race-conditional approval rates $\hat{\beta}$,
\item the adverse-impact ratio under $\hat{\beta}$,
      $\widehat{\text{AIR}}(r)=\hat{\beta}_{r}/\hat{\beta}_{\text{White}}$, for each $r \in \mathcal{R}$
\item the spectral condition number
      $\kappa(\Phi)=\sigma_{\max}(\Phi)/\sigma_{\min}(\Phi)$
      of the posterior race probability matrices, used as a diagnostic for non-singularity.
\end{itemize}
We summarize the empirical estimation error by computing the Root-Mean-Squared Error (RMSE) with respect to the AIR under $\betatrue$.
\[
  \text{RMSE}
    =\sqrt{\frac1K\sum_{k=1}^K
        \bigl(\widehat{\text{AIR}}(r)^{(k)}_{\text{METHOD}}-\widehat{\text{AIR}}(r)_{\text{TRUE}}\bigr)^2}.
\]
Note that $\widehat{\text{AIR}}(r)_{\text{TRUE}}$ is fixed over iterations because it is a function only of the fixed variables rather than the random surname.

\subsection*{Results}

\begin{figure}[H]
  \centering
  \includegraphics[width=0.8\linewidth]{"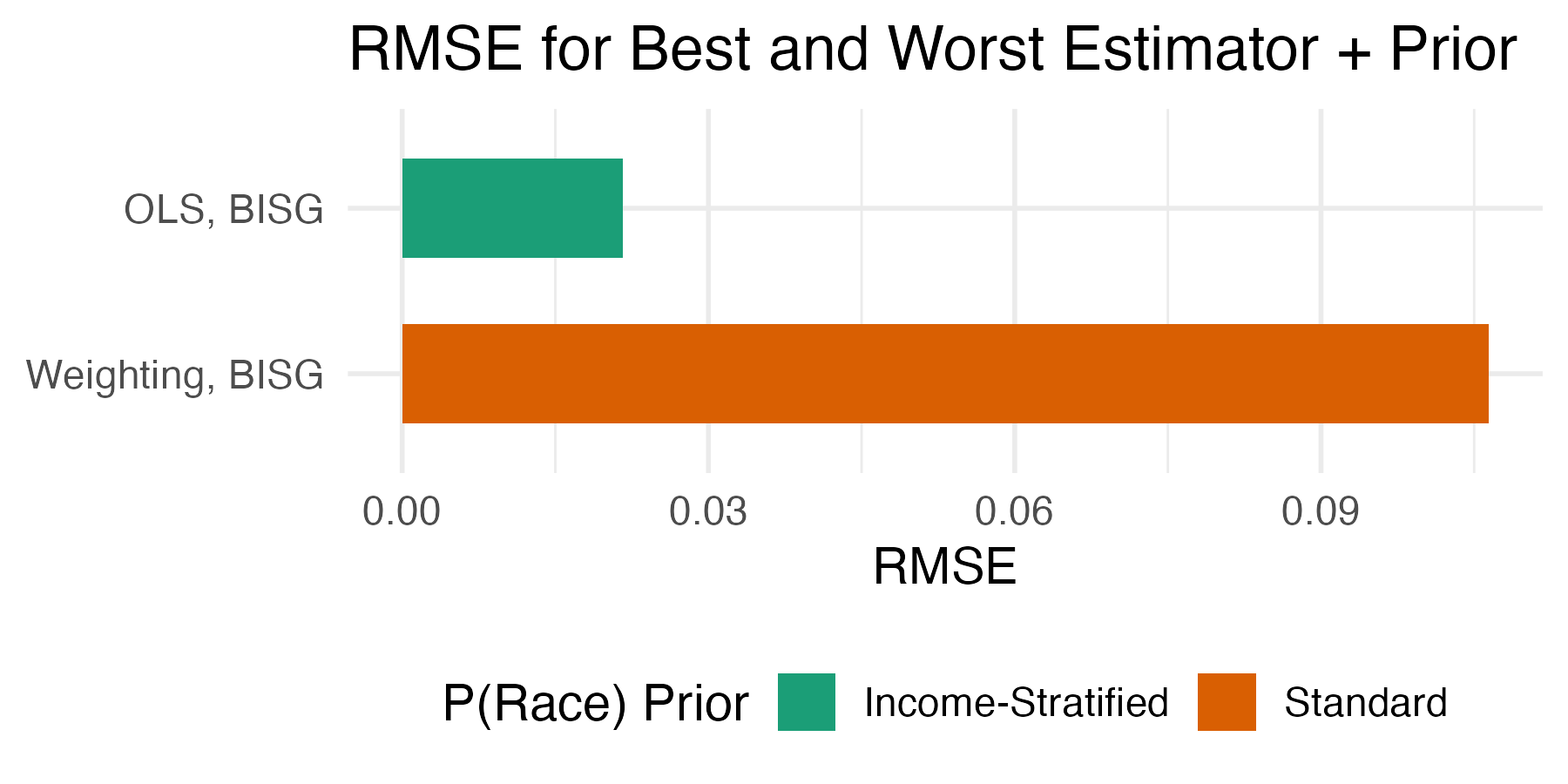"}
  \caption{Best and Worst RMSE for Black/White Adverse Impact Ratio.}
  \label{fig:bestworst}
\end{figure}

The largest observed disparity is between the Black and White groups $\bigl(\widehat{\text{AIR}}(\text{Black})_\text{TRUE}=86.9\%\bigr)$, with others being relatively small $\bigl(\widehat{\text{AIR}}(\text{API})_\text{TRUE}=99.1\% \text{ , }  \widehat{\text{AIR}}(\text{Hispanic})_\text{TRUE}=100.3\%\bigr)$. The OLS method under the income-stratified prior reduces RMSE by $79.7\%$ (from $10.639$\,pp to $2.158$\,pp) relative to the Weighting estimator under the standard prior as shown in \Cref{fig:bestworst}.

\begin{figure}[H]
  \centering
  \includegraphics[width=0.8\linewidth]{"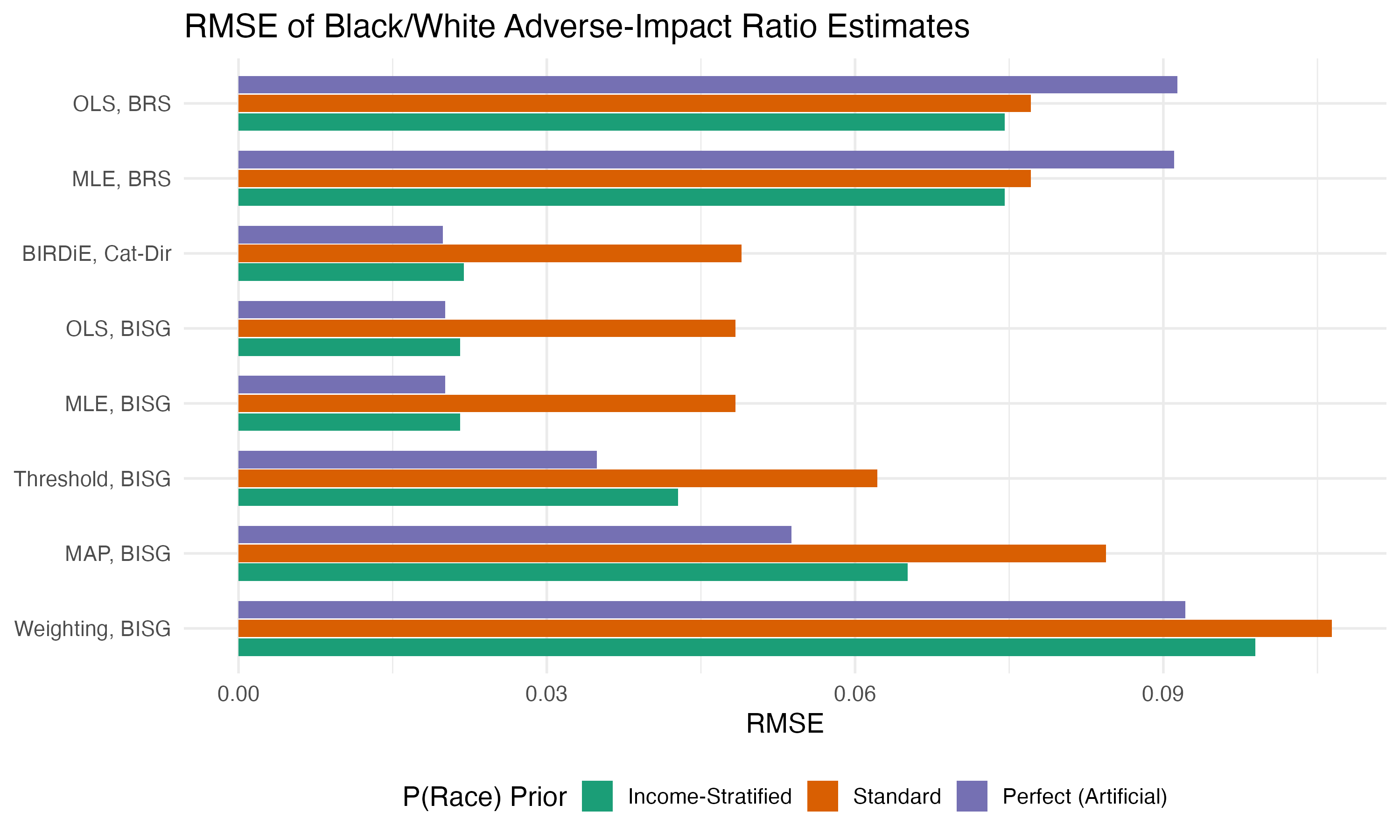"}
  \caption{All RMSEs for Black/White Adverse Impact Ratio.}
  \label{fig:rmse}
\end{figure}

The BISG-based principled estimations exhibit uniformly lower RMSE than the heuristic estimations in estimating $\text{AIR}(\text{Black})$. The principled BISG-based estimators each performed nearly identically, with their best estimates being given under the income-stratified prior. The income-stratified prior makes a marked improvement in reduced empirical estimation error relative to the standard prior. The principled estimations under BRS, while consistent under weaker assumptions, are more sensitive to the random surname and have greater variance (and thus RMSE) over surname Monte Carlo runs. Because the randomization is only with respect to surname in this simulation, the RMSE measured here inherently favors the methods which use fixed geography. The relatively lower condition numbers $\kappa$ under BISG rather than BRS (see \Cref{tbl:kappa}) also indicate greater numerical stability of the estimates under BISG. \Cref{fig:rmse} compares the RMSE of all methods.

\begin{table}[ht]
\centering
\begin{tabular}{llc}
\textbf{Prior} & \textbf{Posterior} & $\kappa$\text{: mean, (95\% CI)}\\
Standard              & BISG & 10.417 (10.243, 10.588)\\
Income‐Stratified     & BISG & 6.229 (6.157, 6.303)\\
Perfect (Artificial)  & BISG & 10.323 (10.168, 10.486)\\[0.25em]
Standard              & BRS  & 29.995 (27.121, 33.063)\\
Income‐Stratified     & BRS  & 18.734 (17.869, 19.611)\\
Perfect (Artificial)  & BRS  & 36.118 (32.717, 39.766)\\
\end{tabular}
\caption{Posterior condition numbers $\kappa$ for each combination of prior and imputation method.}
\label{tbl:kappa}
\end{table}

\begin{figure}[H]
  \centering
  \includegraphics[width=0.8\linewidth]{"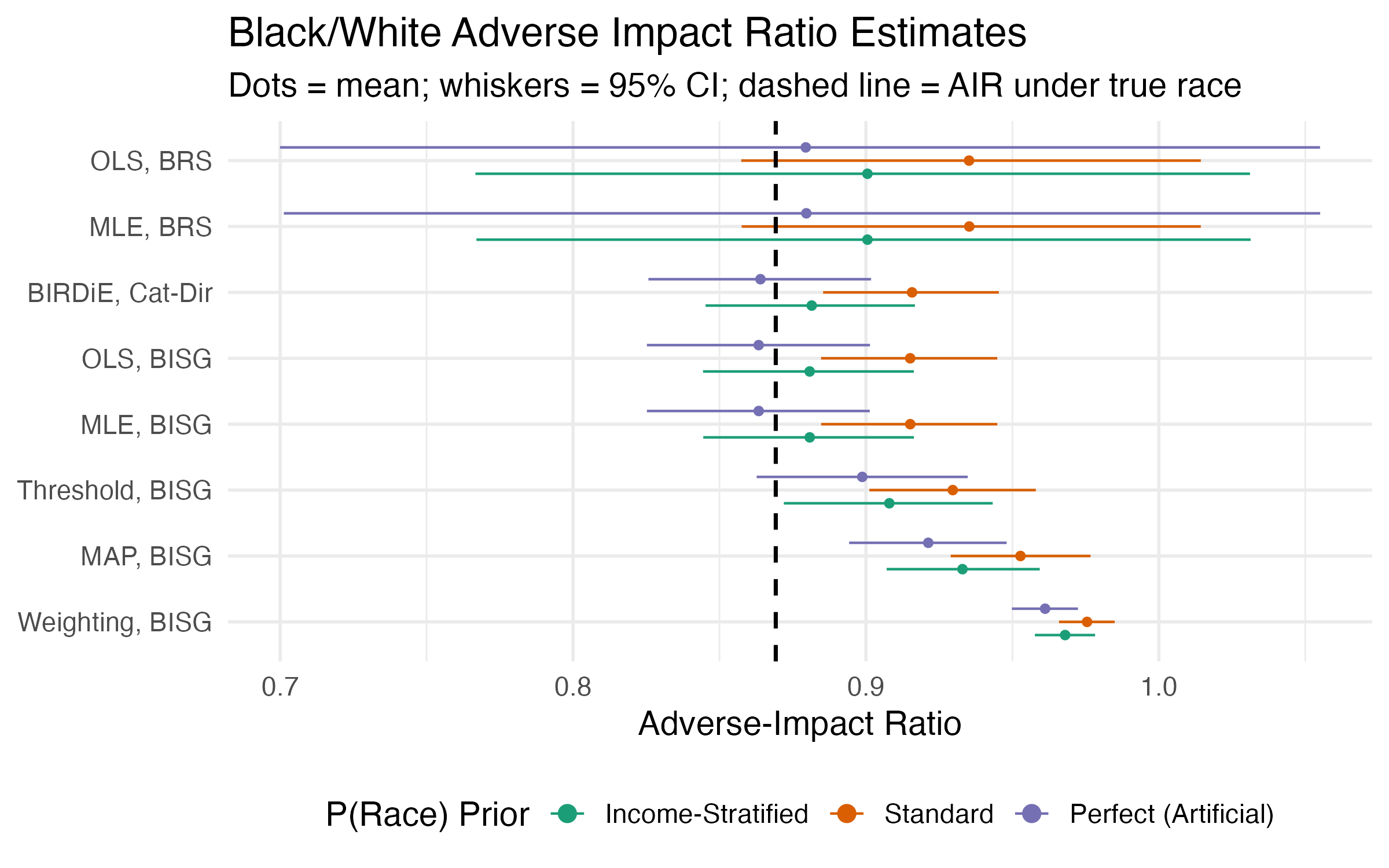"}
  \caption{All Estimates of Black/White Adverse Impact Ratio.}
  \label{fig:AIR}
\end{figure}

\begin{figure}[H]
  \centering
  \includegraphics[width=0.8\linewidth]{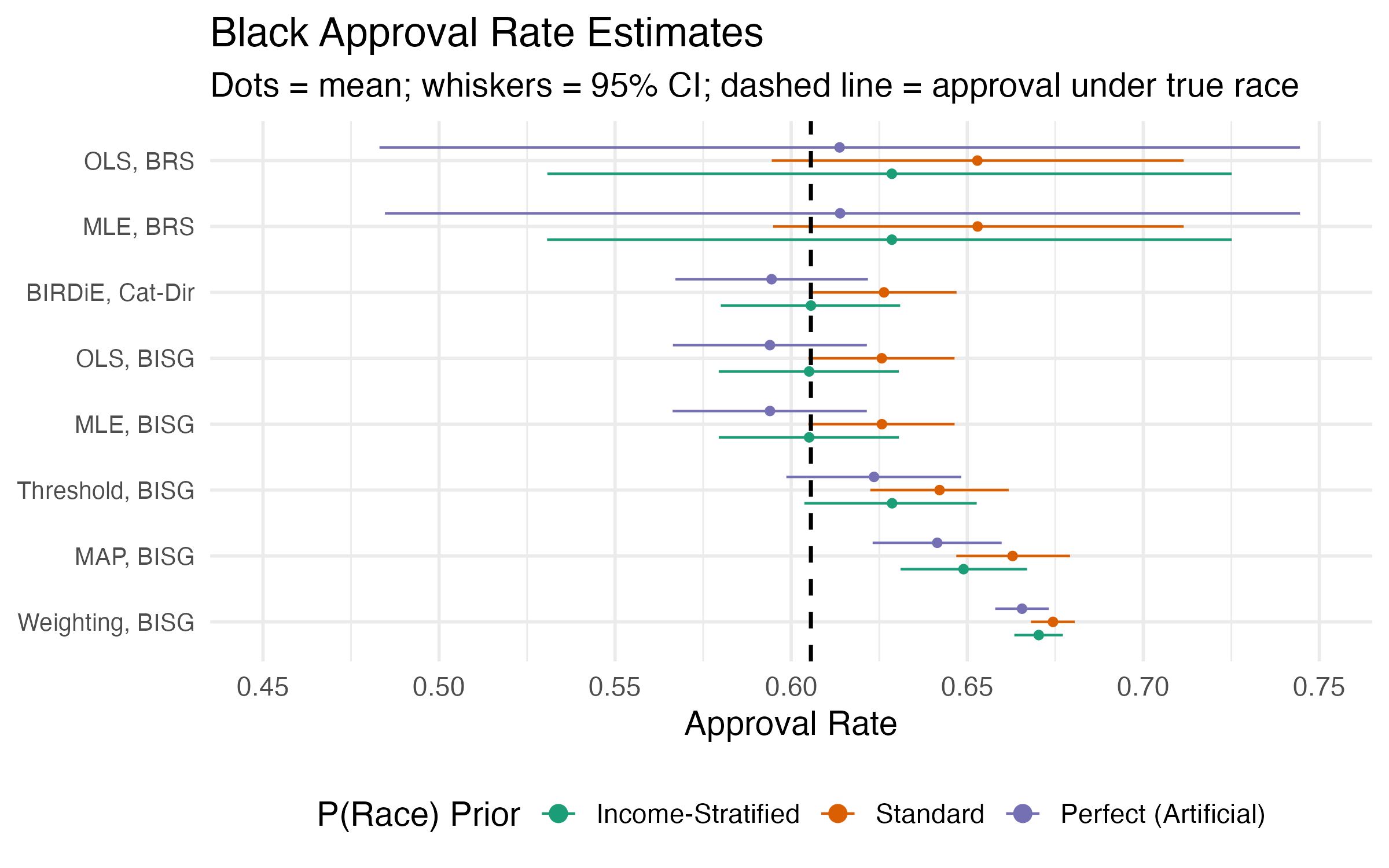}
  \caption{Estimates of Black Approval Rate.}
  \label{fig:BAR}
\end{figure}

\begin{figure}[H]
  \centering
  \includegraphics[width=0.8\linewidth]{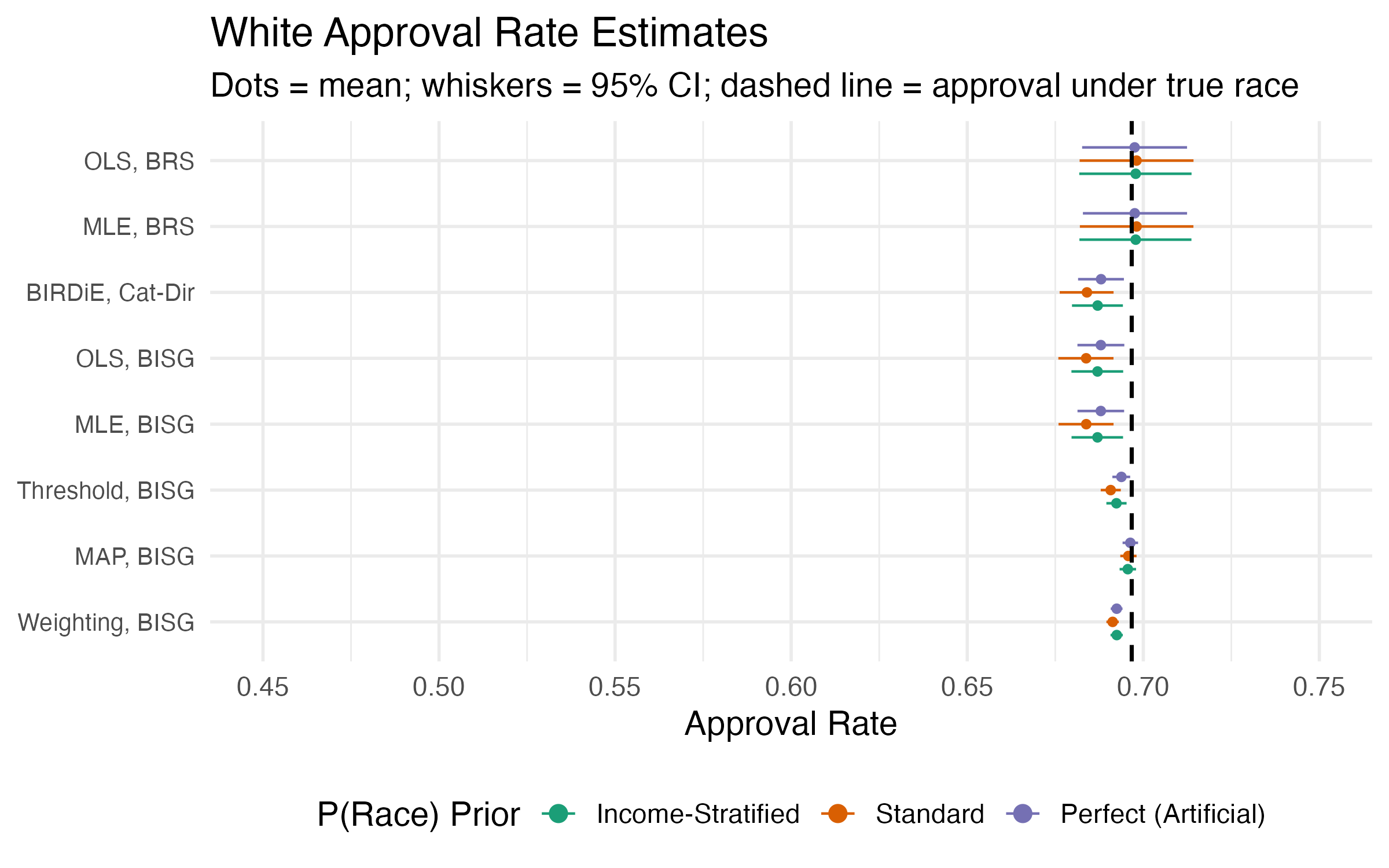}
  \caption{Estimates of White Approval Rate.}
  \label{fig:WAR}
\end{figure}

Critical in fair lending is the sign of the estimation error for the Adverse Impact Ratio. Under-estimation of the AIR results in false positive detection of significant disparities, whilst over-estimation results in false negatives. We find it reasonable to assume that a false negative disparity is more costly to society as well as to the lending institution than a false positive that promptly triggers deeper analysis and potential remediation, though we do not propose exact cost weights. In this case studied here the heuristic estimators each give over-estimates of the AIR, giving an apparent ratio closer to demographic parity than the ratio measured under true race, as we see in \cref{fig:AIR}. We see below in \cref{fig:BAR} and \cref{fig:WAR} that most of this error is due to over-estimation of Black approval rate.

\medskip
The findings on LA HMDA data enriched with synthetic surnames demonstrate that principled estimators with race priors designed to better approximate the applicant population deliver substantial improvements in estimation over heuristic methods in this particular context. It also demonstrates that real-world fair lending applications may violate (ACC) and (CI-YZ), resulting in systematic estimation error. This affirms the importance of assessing the validity of these assumptions though empirical diagnostics where possible as well as domain expertise.

\section*{Conclusion}

We have shown that replacing unobserved race indicators with calibrated posterior race probabilities conditioned on variables whose effect on approval is entirely mediated by race yields a surrogate model that supports both OLS and MLE estimation of race‑conditional approval rates. The calibration of the posterior and conditional independence of the race predictors are strong assumptions which should be carefully tested and considered. When those assumptions are satisfied, the resulting estimators are consistent and asymptotically normal under mild regularity conditions, providing a principled alternative to the heuristic Threshold, MAP, and Weighted‑Average estimators still used widely in fair‑lending practice.  When the geography variable in BISG violates the conditional‑independence assumption, the surname‑only BRS proxy offers a defensible fallback when sample size is sufficient. 

Simulations confirm convergence of the OLS and MLE, highlight the spectral condition number as a critical diagnostic for the full-rank support assumption, and analyze the sensitivity of OLS and MLE to violations of the conditional independence assumption. A semi-synthetic study of 2023 Los Angeles HMDA data demonstrates material differences in measured disparities that could affect regulatory conclusions, with principled estimators providing more accurate results than heuristics. This study also introduces the technique of stratification by income when estimating the prior race distribution to better match the applicant population and meet the calibration assumption.

Future work should investigate (i) robust variance estimation and hypothesis testing for significant disparities, (ii) diagnostics for violation of conditional independence and calibration assumptions, (iii) further methods for improving calibration of posterior race probabilities, (iv) strategies for partial identification when the assumptions of the principled methods cannot be satisfied, and (v) extending the principled methods to continuous outcomes such as price or delay in service. Nonetheless, the strategy for principled estimation studied here is immediately actionable for financial institutions and regulators seeking both statistical soundness and computational ease.
\printbibliography[
  heading=bibintoc,   
  title={References}  
]

\begin{appendices}

\section{Proof of Finite Second Moment \(\Sigma_{gg}\)}
\label{app:proof-second-moment-old}

Here we show that the second moment matrix 
\[
  \Sigma_{gg}\;=\;\mathbb{E}\!\bigl[g_i g_i^{\top}\bigr]
\]
has all finite values.
The finite second moment can be demonstrated by showing that $\mathbb{E}\bigl[\|g_i(\beta)\|^{2}\bigr] = \text{trace}(\mathbb{E}\bigl[g_i(\beta)\bigr]) \leq \infty$. This is because any non-finite covariance term (off-diagonal element) of covariance matrix $\mathbb{E}\bigl[g_ig_i^{\top}\bigr]$ implies at least one non-finite variance term (diagonal element) under Cauchy-Schwarz $$\mathbb{E}\bigl[g_ig_i^{\top}\bigr]_{ab} \leq \mathbb{E}\bigl[g_ig_i^{\top}\bigr]_{aa} \mathbb{E}\bigl[g_ig_i^{\top}\bigr]_{bb}$$ Since the variances are strictly positive, the trace is an upper bound on the values on the diagonal. Any non-finite value in the matrix $\mathbb{E}\bigl[g_ig_i^{\top}\bigr]$ would thus imply a non-finite trace $\mathbb{E}\bigl[\|g_i(\beta)\|^{2}\bigr]$.
\[
\begin{aligned}
\mathbb{E}\bigl[\|g_i(\beta)\|^{2}\bigr] 
   &= \mathbb{E}\bigl[\|\phi_i\|^{2}(Y_i-\phi_i^{\top}\beta)^{2}\bigr] \\[4pt]
   &\le \mathbb{E}\bigl[\|\phi_i\|^{2}2(Y_i^{2}
        + (\phi_i^{\top}\beta)^{2})\bigr] 
        &&\text{(Young’s inequality: }(a-b)^2\le 2(a^2+b^2)\text{)} \\[4pt]
   &= 2\mathbb{E}\bigl[\|\phi_i\|^{2}Y_i^{2}\bigr]
        + 2\mathbb{E}\bigl[\|\phi_i\|^{2}(\phi_i^{\top}\beta)^{2}\bigr] \\[4pt]
   &\le 2\mathbb{E}\bigl[\|\phi_i\|^{2}Y_i^{2}\bigr]
        + 2\mathbb{E}\bigl[\|\phi_i\|^{2}\|\phi_i\|^2 \|\beta\|^{2}\bigr] 
        &&\text{(Cauchy--Schwarz: }(u \cdot v)^2\le\|u\|^{2}\|v\|^{2}\text{)} \\[4pt]
   &\le 2\mathbb{E}\bigl[Y_i^{2}\bigr]
        + 2\mathbb{E}\bigl[\|\beta\|^{2}\bigr] 
        &&\text{because $\|\phi_i\|^{2} = \sum_j^{n}\phi_{ij}^{2} 
        \leq \sum_j^{n}\phi_{ij} = 1$ } \\[4pt]
   &\le 2
        + 2|\mathcal{R}|
        &&\text{because $\mathbb{E}[Y_i]^{2} \leq \mathbb{E}[Y_i] \leq 1$ and $\|\beta\|^{2}\leq|\mathcal{R}|$} \\[4pt]
   &<\infty &&\text{for any finite number of races $|\mathcal{R}|$}
\end{aligned}
\]

\section{Proof of Non-Singularity of \(\Sigma_{gg}\)}
\label{app:proof-id-ols}

\paragraph{Claim.}
Under \textbf{(ID)} and the assumptions that  
\(Y_i\sim\operatorname{Bernoulli}(p)\) with \(0<p<1\) and  
\(g_i=\phi_i\bigl(Y_i-\phi_i^{\top}\beta\bigr)\),  
the matrix  
\[
  \Sigma_{gg}\;=\;\mathbb{E}\!\bigl[g_i g_i^{\top}\bigr]
\]
is nonsingular.

\paragraph{Proof.}
Define  
\(\Sigma_{\phi\phi}=\mathbb{E}[\phi_i\phi_i^{\top}]\)  
and note the factorisation
\[
  \Sigma_{gg}
  \;=\;
  \mathbb{E}\!\Bigl[(Y_i-\phi_i^{\top}\beta)^{2}\,
                    \phi_i\phi_i^{\top}\Bigr].
\]

\medskip\noindent
\emph{Step 1 (singularity transfers).}  
Suppose, for contradiction, that \(\Sigma_{gg}\) is singular.  
Then some non-zero vector \(v\) satisfies
\[
  v^{\top}\Sigma_{gg}v
  \;=\;
  \mathbb{E}\!\bigl[(Y_i-\phi_i^{\top}\beta)^{2}
                    (v^{\top}\phi_i)^{2}\bigr]
  \;=\;0.
\tag{1}
\]
Because the integrand in the expected value function in \((1)\) is non-negative, (1) implies that it must equal zero almost surely.  
Since \(Y_i\) takes the values \(0\) and \(1\) with positive probability and \(\phi_i^{\top}\beta\in(0,1)\), we have \(\mathbb{P}\bigl(Y_i-\phi_i^{\top}\beta\neq 0\bigr)>0\).  
Therefore \((v^{\top}\phi_i)^{2}=0\) almost surely, implying
\[
  v^{\top}\phi_i = 0
  \quad\text{a.s.}
\tag{2}
\]

\medskip\noindent
\emph{Step 2 (consequence for \(\Sigma_{\phi\phi}\)).}  
Taking expectations in \((2)\) gives
\[
  v^{\top}\Sigma_{\phi\phi}v
  =\mathbb{E}\!\bigl[(v^{\top}\phi_i)^{2}\bigr]
  =0,
\]
so \(\Sigma_{\phi\phi}\) would be singular.

\medskip\noindent
\emph{Step 3 (contradiction with full-rank support).}  
Assumption \textbf{(ID)} states that \(\Sigma_{\phi\phi}\) is non-singular.  
This contradicts the conclusion of Step 2, so the initial supposition must be false.  
Hence \(\Sigma_{gg}\) cannot be singular.

\hfill\(\square\)

\section{Derivation for Score, Hessian, and Concavity of the Surrogate Log-Likelihood}
\label{app:proof-concave-mle}

Let
\[
  p_i \;=\; \phi_i^{\!\top}\theta,
  \qquad
  \ell_i(\theta)
  \;=\;
  Y_i\log(p_i) \;+\; (1-Y_i)\log\!\bigl(1-p_i\bigr).
\]

\paragraph{1. Differentiate w.r.t.\ the Bernoulli parameter \(p_i\).}
\[
  \frac{\partial\ell_i}{\partial p_i}
  \;=\;
  \frac{Y_i}{p_i}
  \;-\;
  \frac{1-Y_i}{1-p_i}.
  \tag{A}
\]

\paragraph{2. Apply the chain rule to get the gradient w.r.t.\ \(\theta\).}
Because \(p_i=\phi_i^{\!\top}\theta\),
\[
  \frac{\partial p_i}{\partial\theta}
  \;=\;
  \phi_i.
  \tag{B}
\]
Hence
\[
  \nabla_\beta\ell_i(\theta)
  \;=\;
  \frac{\partial\ell_i}{\partial p_i}\;
  \frac{\partial p_i}{\partial\theta}
  \;=\;
  \phi_i\!
  \Bigl[
      \frac{Y_i}{p_i}
      - \frac{1-Y_i}{1-p_i}
  \Bigr].
\]

\paragraph{3. Stack over \(i=1,\dots,n\).}
Define \(p=\Phi\theta\) (element-wise), then
\[
  \boxed{\;
      \nabla_\beta\ell(\beta)
      = \Phi^{\!\top}
        \bigl[
            Y\,/\,p
            - (1-Y)\,/\,\bigl(1-p\bigr)
        \bigr]
  \;}
\]
where “/” denotes element-wise division.

\paragraph{4. Hessian.}
Let
\[
  d \;=\; Y\,/\,p^{2} + (1-Y)\,/\,\bigl(1-p\bigr)^{2},
  \quad\text{(element-wise)}
\]
then
\[
  \boxed{\;
      \nabla_\theta^{2}\ell(\theta)
      = -\,\Phi^{\!\top}\operatorname{diag}(d)\,\Phi
  \;}
\]
and \(-\nabla^{2}\ell\) is the observed information matrix.

\paragraph{5. Concavity.}
For any \(v\in\mathbb{R}^{p}\),
\[
  v^{\!\top}\nabla_\beta^{2}\ell(\beta)\,v
  \;=\;
  -\sum_{i=1}^{n} d_i\,(\phi_i^{\!\top}v)^{2}
  \;\le\;0,
\]
because each \(d_i>0\).  
Hence \(\nabla_\theta^{2}\ell(\beta)\) is negative-semidefinite on the domain \(0<\Phi\theta<1\); therefore the log-likelihood \(\ell(\theta)\) is concave.  Consequently, the observed information matrix  
\(-\nabla_\theta^{2}\ell(\theta)=\Phi^{\!\top}\operatorname{diag}(d)\Phi\) is positive-semidefinite. 

\section{Proof of Non-Singularity of the Information Matrix \(I(\beta)\)}
\label{app:proof-id-mle}

\paragraph{Claim.}
Under the stated assumptions and surrogate model,
the Fisher information at the true parameter
\[
  I(\beta)\;=\;\mathbb{E}\!\bigl[-\nabla_{\beta}^{2}\ell_i(\beta)\bigr]
\]
is nonsingular.

\paragraph{Proof.}
For a single observation the Hessian is  
\[
  \nabla_{\theta}^{2}\ell_i(\theta)
  =-\phi_i\phi_i^{\!\top}\,d_i(\theta),
  \qquad
  d_i(\theta)=
    \frac{Y_i}{\bigl(\phi_i^{\!\top}\theta\bigr)^{2}}
    +\frac{1-Y_i}{\bigl(1-\phi_i^{\!\top}\theta\bigr)^{2}} .
\]

\medskip\noindent
\emph{Step 1 (assume \(I(\beta)\) singular).}  
Suppose, for contradiction, that \(I(\beta)\) is singular.  
Then there exists \(v\neq0\) such that
\[
  v^{\top}I(\beta)v
  \;=\;
  \mathbb{E}\!\bigl[d_i(\beta)\,(v^{\top}\phi_i)^{2}\bigr]
  \;=\;0.
  \tag{1}
\]

\medskip\noindent
\emph{Step 2 (zero dot product).}  
Because \(d_i(\beta)>0\) almost surely (INT and \(Y_i\in\{0,1\}\)),
equation (1) forces \((v^{\top}\phi_i)^{2}=0\) almost surely, i.e.  
\[
  v^{\top}\phi_i=0
  \quad\text{a.s.}
  \tag{2}
\]

\medskip\noindent
\emph{Step 3 (implication for \(\Sigma_{\phi\phi}\)).}  
Taking expectations in (2) gives
\[
  v^{\top}\Sigma_{\phi\phi}v
  =\mathbb{E}\!\bigl[(v^{\top}\phi_i)^{2}\bigr]
  =0,
\]
so \(\Sigma_{\phi\phi}:=\mathbb{E}[\phi_i\phi_i^{\!\top}]\) would be singular.

\medskip\noindent
\emph{Step 4 (contradiction).}  
Assumption \textbf{(ID)} asserts that \(\Sigma_{\phi\phi}\) is nonsingular.  
This contradicts Step 3, so the initial supposition is false.  
Hence \(I(\beta)\) must be nonsingular.

\hfill\(\square\)

\section{Simulation Algorithm}
\label{app:sim}

This appendix explains precisely how one Monte--Carlo iteration produces  
(i) a true‐race indicator matrix \(T\),  
(ii) a BISG posterior probability matrix \(\Phi^{\text{BISG}}\), and  
(iii) a loan–approval vector \(Y\)  
for a sample of \(n\) credit applications.  
Throughout we fix five racial categories, five surnames, and five geographic
areas; all notation below inherits that convention.

\subsection{General four–step procedure}

\paragraph{Step 1: Draw true race.}

Each application \(i\in\{1,\dots,n\}\) independently receives a race
\(R_i\in\mathcal R=\{\text{R1},\dots,\text{R5}\}\) from the
common prior
\(\mathbb{P}(R_i=r)=1/5.\)
Stacking the one–hot vectors
\(t_i\)
gives the \(n\times5\) indicator matrix
\(T=[t_1^{\!\top}\,\dots\,t_n^{\!\top}]^{\!\top}.\)

\paragraph{Step 2: Generate surname and geography.}

For each race \(r\) we fix two conditional distributions  

\[
\mathbb{P}(S=s\mid R=r) \in \Delta^4,
\qquad
\mathbb{P}(G=g\mid R=r) \in \Delta^4,
\]

\noindent collected as \(5\times5\) stochastic matrices
\(\mathcal{S}\) (surname) and \(\Gamma\) (geography).  
Given \(R_i=r\), the applicant’s surname \(S_i\) and geography \(G_i\)
are sampled independently from the corresponding rows of
\(\mathcal{S}\) and \(\Gamma.\) We write $\Delta^4$ for the standard 4-simplex.

\paragraph{Step 3: Compute BISG posteriors.}

For each applicant the BISG probability that the race corresponding with the loan application is \(r\) is  

\[
\Phi^{\text{BISG}}_{ir}
  \;=\;
  \frac{\mathbb{P}(S_i\mid R=r)\,
        \mathbb{P}(G_i\mid R=r)\,
        \mathbb{P}(R=r)}
       {\sum_{r'}\mathbb{P}(S_i\mid R=r')\mathbb{P}(G_i\mid R=r')\mathbb{P}(R=r')},
  \qquad r=1,\ldots,5.
\]

\noindent We assemble these rows to form the \(n\times5\) matrix
\(\Phi^{\text{BISG}}\).

\paragraph{Step 4: Generate approval outcomes.}

Let the baseline race–conditional approval probabilities be the fixed vector  

\[
\beta=(0.80,\;0.50,\;0.60,\;0.20,\;0.55).
\]

\noindent Each approval indicator is drawn as  

\[
Y_i\;\sim\;\text{Bernoulli}\bigl(p_i\bigr),
\qquad
p_i=\beta_{R_i}+\alpha\,\mathbf1\{G_i=\text{G5}\},
\]

\noindent where \(\alpha\) is an additive geography–five effect
used only in the \textit{CI–YZ} scenario and clipped to keep \(0<p_i<1\). Stacking the \(Y_i\) gives the length–\(n\) vector \(Y\). After these four steps the data for that iteration  \(\bigl(T,\Phi^{\text{BISG}},Y\bigr)\) have been generated.

\subsection{Scenario–specific choices for \(\mathcal{S}\), \(\Gamma\) and \(\alpha\)}

Let \(\xi>0\) denote a localisation tuning constant and
\(\mathrm{Dir}_5(\mathbf 1)\) the five–dimensional Dirichlet distribution
that is uniform on the \(4\)-simplex.
Rows are indexed by race \(r\in\{1,\dots,5\}\) and columns by surname
\(s\) or geography \(g\).

\begin{description}
\item[True‐label (\textbf{A}).]  

\[
  \mathcal S_{rs}= \mathbf 1\{r=s\},\qquad
  \Gamma_{rg}= \mathbf 1\{r=g\},\qquad
  \alpha=0.
\]

\textit{Interpretation.}  
Every race deterministically maps to a unique surname and geography,
so the BISG posterior collapses to a one–hot matrix:
information is \emph{perfect} and entropy is zero.

\item[Localised probability (\textbf{B}).]  

Draw i.i.d.\ jitters \(\varepsilon_{rs}\sim\mathrm{Unif}(0,\xi)\) with
\(\xi=1\) and set  

\[
  \tilde{\mathcal S}_{rs}= \mathbf 1\{r=s\}+\varepsilon_{rs},\qquad
  \mathcal S_{rs}= \frac{\tilde{\mathcal S}_{rs}}
                        {\sum_{s'}\tilde{\mathcal S}_{rs'}}
\]
with the same formula applied for \(\Gamma\); set \(\alpha=0\).

\textit{Interpretation.}  
Each row still strongly favours its diagonal category, but the small
jitter injects mild ambiguity.  Posterior rows are sharply peaked
(\emph{high-information}) yet not degenerate, mirroring the behaviour of
real-world BISG scores when either surname or geography is distinctive.

\item[Uniform probability (\textbf{C}).]  

Sample independently  

\[
  (\mathcal S_{r1},\dots,\mathcal S_{r5})
  \sim\mathrm{Dir}_{5}(\mathbf 1),\qquad
  (\Gamma_{r1},\dots,\Gamma_{r5})
  \sim\mathrm{Dir}_{5}(\mathbf 1),
\]
and set \(\alpha=0\).

\textit{Interpretation.}  
Rows are spread nearly uniformly across all five categories, giving
\emph{low-information, high-entropy} posteriors—a harder case for estimation than the localized probabilities.

\item[CI–YZ violation (\textbf{D}).]  

Reuse the \textbf{B} localisation with \(\xi=1\) but introduce a direct
geography effect  

\[
  \alpha \in (-0.55,\;0.45)
\]
added only when \(G_i=\text{G5}\).

\textit{Interpretation.}  
The BISG probabilities are still informative, but approval now depends
on geography beyond its correlation with race, deliberately breaking the
assumption \(Y\!\perp\! G \mid R\).  This stresses the principled
estimators and reveals their sensitivity to mis-specification.
\end{description}

\noindent These variations are the only differences across scenarios; all remaining
sampling steps are identical. Each Monte–Carlo repetition therefore
yields a fresh data set reflecting the chosen structural assumptions.

\section{Derivation of Total Variation Distance}
\label{app:proof-tvd}

We define the joint probability measure for $(R,Y)$ under a given race-conditional approval probability vector $u$ as
\begin{align*}
M(u) \mapsto P(r,y) &= 
\mathbb{P}(R) \cdot
  \begin{cases}
     \mathbb{P}(Y \mid R),        & y=1,\\[2pt]
     1-\mathbb{P}(Y \mid R),      & y=0,
  \end{cases} \\
&=
\frac{1}{5} \cdot
  \begin{cases}
     u_r,        & y=1,\\[2pt]
     1-u_r,      & y=0,
  \end{cases}
\end{align*}

\noindent We then derive the total variation distance as follows:

\begin{align*}
\mathrm{TVD}(M(u),M(v))
  &:=\sup_{A\subseteq\mathcal R\times\{0,1\}}
      \lvert M(u)-M(v)\rvert  \\[3pt]
  &=\frac12\sum_{r=1}^{5}\sum_{y\in\{0,1\}}
      \lvert M(u)-M(v)\rvert \\[3pt]
  &=\frac12\sum_{r=1}^{5}\frac{1}{5}
      \Bigl(\lvert u_r-v_r\rvert
            +\lvert(1-u_r)-(1-v_r)\rvert\Bigr) \\[3pt]
  &=\frac{1}{5}\sum_{r=1}^{5}\lvert u_r-v_r\rvert
      \qquad\text{since }\lvert(1-a)-(1-b)\rvert=\lvert a-b\rvert. \\[6pt]
\end{align*}

\end{appendices}

\end{document}